\documentclass[a4paper,12pt]{article}

\usepackage{amsmath}
\usepackage[psamsfonts]{amssymb}
\usepackage{rsfs}
\usepackage{bm}

\usepackage{cite}
\usepackage[dvips]{graphicx}

\makeatletter
\@addtoreset{equation}{section}
\makeatother

\begin{document} 

\begin{titlepage}

\baselineskip 10pt
\hrule 
\vskip 5pt
\leftline{}
\leftline{Chiba Univ. Preprint
          \hfill   \small \hbox{\bf CHIBA-EP-135}}
\leftline{\hfill   \small \hbox{hep-th/0206173}}
\leftline{\hfill   \small \hbox{June 2002}}
\vskip 5pt
\baselineskip 14pt
\hrule 
\vskip 1.0cm
\centerline{\large\bf 
A Confining String Theory 
} 
\vskip 0.5cm
\centerline{\large\bf  
Derivable from Yang-Mills Theory 
}
\vskip 0.5cm
\centerline{\large\bf  
due to a Novel Vacuum Condensate 
}
\vskip 0.5cm
\centerline{\large\bf  
}

\vskip 0.5cm

\centerline{{\bf 
Kei-Ichi Kondo$^{\dagger,{\ddagger},{1}}$ and Takahito Imai$^{\ddagger,{2}}$
}}  
\vskip 1cm
\begin{description}
\item[]{\it \centerline{  
${}^{\dagger}$Department of Physics, Faculty of Science, 
Chiba University,  Chiba 263-8522, Japan}
  }
\item[]{\it 
${}^{\ddagger}$Graduate School of Science and Technology, 
Chiba University, Chiba 263-8522, Japan
  }
\end{description}

\begin{abstract}
We discuss within the weak-field approximation and the derivative expansion how the area law of the Wilson loop follows directly from the vacuum condensate of mass dimension 2, i.e., simultaneous Bose-Einstein condensation of gluon pair and ghost-antighost pair.
Such a novel vacuum condensate was recently claimed to exist as the non-vanishing vacuum  expectation value of a BRST-invariant composite operator of mass dimension 2.  
First of all, we use a version of the non-Abelian Stokes theorem to rewrite the Wilson loop line integral to a surface integral.  Then we convert the Yang-Mills theory with an insertion of the Wilson loop operator into a bosonic string theory with a rigidity term by way of an equivalent antisymmetric tensor gauge theory which couples to the surface spanned by the Wilson loop.   
This result suggests an intimate relationship between quark confinement and mass gap in Yang-Mills theory.
In fact, the dual Ginzburg-Landau theory describing the dual superconductivity is also derivable by making use of duality transformations without using the naive Abelian projection and without breaking the global color invariance of the original Yang-Mills theory.  This feature is desirable from the viewpoint of color confinement preserving color symmetry.

\end{abstract}


\vskip 0.2cm
\hrule  
\vskip 0.2cm
${}^1$ 
  E-mail:  {\tt kondo@cuphd.nd.chiba-u.ac.jp}

${}^2$ 
  E-mail:  {\tt takahito@physics.s.chiba-u.ac.jp}

\vskip 0.2cm  

\par 
\par\noindent
\vskip 0.5cm


\vskip 0.5cm

\newpage
\pagenumbering{roman}
\tableofcontents

\vskip 0.5cm  



\end{titlepage}


\pagenumbering{arabic}

\baselineskip 14pt
\section{Introduction}

Proving quark confinement is still one of the most important problems in theoretical physics.
Based on intensive investigations in the last decade, 
the dual superconductor picture \cite{dualsuper} for the vacuum of quantum chromodynamics (QCD) is believed to be the most promising scenario of deriving quark confinement.
To derive this picture, it is convenient to select out an Abelian sector of the original Yang-Mills theory so that the duality transformation of the Abelian sector gives a dual Abelian gauge theory in which the dual superconductivity is caused by magnetic monopole condensation.  Such a procedure is called the Abelian projection \cite{tHooft81}.  Therefore, the Abelian projected effective gauge theory of the original Yang-Mills theory has been intensively studied in the last decade from the viewpoint of 
numerical \cite{KLSW87,SY90,SNW94,Miyamura95,review} and  analytical \cite{EI82,Suzuki88,KondoI} methods.
 
\par
Nevertheless, we wish to point out in this paper that {\it the naive procedure of the Abelian projection is not necessarily indispensable to derive the dual superconductivity} for explaining quark confinement.  
Even if we begin with the Lorentz gauge $\partial_\mu \mathscr{A}_\mu(x)=0$ which has {\it global gauge invariance}, the dual superconductivity can be derived without breaking the original global gauge symmetry, as will be demonstrated in this paper.  This result should be compared with the Maximal Abelian (MA) gauge which breaks the global gauge symmetry from the original gauge group to the maximal torus group (except for the discrete Weyl symmetry \cite{IS00}).
This choice of Lorentz gauge can greatly simplify the actual calculations by virtue of preserving the global gauge invariance.
This fact is also desired for enabling us to tackle color confinement problem (i.e., confinement of all color non-singlet objects) beyond quark confinement problem.  
We want to add that we do not deny the MA gauge as a choice of gauge for deriving quark confinement, since quark confinement should be a gauge invariant phenomenon.  
Thus we conclude that the dual superconductivity of QCD vacuum is an intrinsic property, independent of the choice of the Abelian projection to define the monopoles.  
\par
In this paper we adopt the Becchi-Rouet-Stora-Tyutin \cite{BRST} (BRST) invariance as the first principle to characterize the {\it quantized} Yang-Mills theory, rather than the gauge invariance which is broken by the procedure of gauge fixing in the course of quantization. 
The main purpose of this paper is to demonstrate how to derive a confining bosonic string representation of Yang-Mills theory.
In this derivation, we show that {\it the Wilson loop average exhibits the area law decay if the vacuum condensation 
 of mass dimension 2 occurs in Yang-Mills theory}, i.e., the vacuum expectation value (VEV) of the composite operator \cite{Kondo01}, 
\begin{equation}
   \mathcal{O}  
  = \Omega^{-1} \int d^4x \ {\rm tr}_{G/H} \left[ {1 \over 2} \mathscr{A}^\mu(x) \mathscr{A}_\mu(x) + \lambda i \bar{\mathscr{C}}(x) \mathscr{C}(x) \right] ,
\label{op}
\end{equation}
is non-vanishing
$\langle \mathcal{O} \rangle_{YM} \not=0$, 
where $\Omega$ is the volume of the space-time $\Omega:=\int d^4x$
and $\lambda$ is a gauge fixing parameter in the Lorentz gauge $\partial^\mu \mathscr{A}_\mu(x)=0$.  
\par
The vacuum condensates of mass dimension 2 were recently proposed by several authors 
\cite{Schaden99,KS00,Boucaudetal00,GSZ01} and the physical implications have been extensively studied in a couple of years 
\cite{GZ01,Boucaudetal01,Boucaudetal02,Kondo01}. 
Recently, it has been shown \cite{Kondo01} that the composite operator $\mathcal{O}$ of mass dimension 2 can be made both  BRST and anti-BRST invariant%
\footnote{Here the BRST should be understood as the on-shell version (\ref{BRST2}) of the BRST transformation which is obtained by eliminating the  (Nakanishi-Lautrup) auxiliary field $\mathscr{B}$ from the off-shell BRST transformation (\ref{BRST1}).
Note that the operator $\mathcal{O}$ does not include the $\mathscr{B}$ field.
}
in the generalized Lorentz gauge \cite{CF76} and the modified MA gauge \cite{KondoII}.  
Especially, in the limit of Landau gauge $\lambda \rightarrow 0$,
the vacuum expectation value of $\mathcal{O}$ reduces to the gluon condensation of mass dimension 2 proposed in \cite{LS88}, although the way of taking the limit is not unique, see \cite{KMSI02}.
Therefore, we claim in this paper that {\it quark confinement follows from the vacuum condensate of mass dimension 2} in a BRST and anti-BRST invariant manner, as suggested in the previous paper \cite{Kondo01}.  

\par
In this paper, we emphasize the importance of a novel gluon condensate of mass dimension 2, in contrast with the conventional gluon condensate of mass dimension 4, i.e., 
$
 \langle \mathcal{F}_{\mu\nu}^2 \rangle 
$
which is obviously gauge invariant 
 \cite{SVZ79} and BRST invariant.%
\footnote{The derivation of a confining string theory due to gluon condensate of mass dimension 4 has been already tried by several authors, see e.g. a review by Antonov \cite{Antonov99} and references therein.
}
In fact, our results indicate an intimate connection between the existence of mass gap and quark confinement in Yang-Mills theory, since the non-vanishing vacuum condensate of mass dimension 2 leads to  the dynamical generation of the effective masses of gluon and ghost.  
In order to elucidate this statement, we rewrite the Yang-Mills theory in the presence of a Wilson loop operator into a bosonic string theory within the weak-field approximation.  It turns out that the string action is given by a sum of the Nambu-Goto term and the rigidity term (as announced in \cite{Kondo01proc}).  Therefore, the resulting string is called the rigid string hereafter, following  Polyakov \cite{Polyakov86,Polyakov87}, see also
\cite{Kleinert86,BPT87,OY87,BZ87,DG88,David89,PY92,PWZ93,Lee93,Orland94,SY95,KC96}.  

\par
We do not attempt to analyze in detail the physics of the obtained  rigid string, since such investigations have been performed by numerous authors in recent years. 
See e.g., 
\cite{Polchinski92,tHooft74a,KR74,oldstring,LSW80,Alvarez81,Arvis83,Olesen85} 
for the old works of string representation of QCD
and  
\cite{Polyakov96,DQT96,DT97,Polyakov97,AGO98,Ellwanger98,AE98,AE99,Parthasarathy99,Zakharov99,BS00,Kondo00} 
for recent developments on this subject. 
The novel result of this paper is the determination of the parameters in the rigid string action in terms of the original Yang-Mills theory (within the approximations adopted).  Therefore, their result can also be applied  to our string theory by substituting the determined parameters in our paper into their results without repeating the calculations. 
The main purpose of this work is to build bridges between Yang-Mills theory and a rigid string theory based on the existence of a novel vacuum condensate.  
We show that an antisymmetric tensor gauge theory (ASTGT) of rank 2 plays the dominant role in establishing the relationship, as expected in the previous investigations. 

\par
Our analysis begins with a recent observation that non-perturbative corrections to perturbative results do indeed exist even in the high-energy region.   
Such corrections can lead for example to the linear static potential for a pair of quark and anti-quark at {\it short} distance, although the linear static potential is expected to occur in the long distance according to the conventional wisdom.%
\footnote{
Therefore, our study on quark confinement does not imply the existence of $1/p^4$ {\it infrared} singularity in the gluon propagator which was investigated e.g. by Ellwanger \cite{Ellwanger98}.  
}
Consequently, a {\it short} confining string will be obtained as the string representation of the Yang-Mills theory in the relevant energy region (except for the extremely low-energy). 
Extremely low-energy region is still beyond our approach due to lack of suitable tools of analysis. 
We hope that the strategy given in this paper will survive in the low-energy region, if the suitable tools for the calculation become available.  
\par
This paper aims to demonstrate that quark confinement can be understood at least qualitatively even in the lower order of approximations (expansions) adopted, just as the area decay law of the Wilson loop average can be derived even in the lowest order of the strong coupling expansion in lattice gauge theory.
In lattice gauge theory, it is necessary to control the continuum limit for the complete proof of quark confinement.  
In the continuum formulation, this problem is absent, but we must overcome other difficult problems to remove the approximations.  
\par
The outline of this paper is as follows.  
In section 2, we enumerate the necessary steps to arrive at the main results.  In section 3, we give some remarks on the results obtained in section 2.  
In the final section, we give conclusion and discussion.  In order to clarify the essence of this work, technical details are all omitted in section 2.  They are given in Appendices A to F. 

\section{Steps to arrive at the main results}

The main results of this paper are obtained following the steps enumerated below.  
\par
\subsection{Step 1: Definition of the Wilson loop operator and its average}
The {\it Wilson loop operator} $W_C[\mathscr{A}]$ is defined by the trace of the path-ordered product of the line integral of the non-Abelian gauge field $\mathscr{A}_\mu$ i.e., Lie-algebra $\mathcal{G}$ valued one-form $\mathscr{A} = \mathscr{A}_\mu dx^\mu = \mathscr{A}_\mu^A T^A dx^\mu$, along a closed loop $C$:
\begin{equation}
  W_C[\mathscr{A}] = \mathcal{N}^{-1} {\rm tr} \left\{ \mathcal{P} \exp \left[ i g \oint_{C} dx^\mu \mathscr{A}_\mu(x) \right] \right\} ,
\end{equation}
where $\mathcal{N}$ is the normalization factor to guarantee $W_C[0]=1$ and we adopt the convention in which the coupling constant $g$ is explicitly written (For a precise definition of the path ordering $\mathcal{P}$, see e.g., \cite{KT99}).
 In what follows, we assume that the loop $C$ is a connected closed path without self-intersections.  
 \par
The {\it Wilson loop average} $W(C)$ is defined as the vacuum expectation value (VEV) of the Wilson loop operator $W_C[\mathscr{A}]$  in the Yang-Mills theory.  
For concreteness, we suppose that the manifestly Lorentz covariant formulation of the Yang-Mills theory is adopted.  
In this paper, we always work in the Euclidean formulation, unless otherwise stated.  
In the functional integral formulation, the Wilson loop average is defined by
\begin{align}
  W(C) = \langle W_C[\mathscr{A}] \rangle_{YM} 
  = Z_{YM}^{-1} \int d\mu_{YM} e^{-S_{YM}^{tot}} W_C[\mathscr{A}] ,
\end{align}
where $d\mu_{YM}$ is the measure of functional integration, 
\begin{equation}
d\mu_{YM}=\mathcal{D}\mathscr{A}_\mu \mathcal{D}\mathscr{B} \mathcal{D}\mathscr{C} \mathcal{D}\bar{\mathscr{C}}
\end{equation}
defined as the product measure of the respective field, i.e., the gauge (gluon) field $\mathscr{A}_\mu$, Nakanishi-Lautrup (NL) auxiliary field $\mathscr{B}$,  Faddeev-Popov (FP) ghost field $\mathscr{C}$ and antighost field $\mathscr{\bar{C}}$,
and $S_{YM}^{tot}$ is the total action of the Yang-Mills theory consisting of the pure Yang-Mills term $S_{YM}$ for a gauge group $G$, the gauge-fixing (GF) term $S_{GF}$ and the associated FP ghost term $S_{FP}$, i.e.,
$S_{YM}^{tot}=S_{YM}+S_{GF}+S_{FP}$.
The gauge-fixing part $S_{GF+FP}:=S_{GF}+S_{FP}$ is determined based on the Becchi-Rouet-Stora-Tyutin \cite{BRST} (BRST) transformation 
$\bm{\delta}_{\rm B}$.  The BRST transformation in Euclidean space is given by
\begin{subequations}
\begin{align}
 \bm{\delta}_{\rm B} \mathscr{A}_\mu(x)
   & =\mathscr{D}_\mu[\mathscr{A}]\mathscr{C}(x)
    :=\partial_\mu \mathscr{C}(x)
      + g (\mathscr{A}_\mu(x) \times \mathscr{C}(x)) , \\
 \bm{\delta}_{\rm B} \mathscr{C}(x)
   & =-{1 \over 2}g(\mathscr{C}(x) \times \mathscr{C}(x)) , \\
 \bm{\delta}_{\rm B} \bar{\mathscr{C}}(x)
   & = - \mathscr{B}(x) , \\
 \bm{\delta}_{\rm B} \mathscr{B}(x)
   &=0 .
\end{align}
\label{BRST1}
\end{subequations}
In fact, the gauge-fixing part $S_{GF+FP}:=S_{GF}+S_{FP}$ is written in the BRST exact form \cite{KU82}, i.e., $S_{GF+FP} = \bm{\delta}_{\rm B}(\cdots)$ using a quantity $(\cdots)$ depending on the gauge-fixing condition.   
Therefore, the explicit form of $S_{GF+FP}$ depends on the gauge fixing condition.
The pure Yang-Mills action is BRST invariant, $\bm{\delta}_{\rm B}S_{YM}=0$.
The gauge-fixing part is BRST invariant,  
$\bm{\delta}_{\rm B}S_{GF+FP}=0$ due to nilpotency of the BRST transformation $\bm{\delta}_{\rm B}^2 \equiv 0$.  Consequently, the total Yang-Mills action is BRST invariant, 
$\bm{\delta}_{\rm B}S_{YM}^{tot}=0$.
\par
For our purposes, however, it is more convenient to eliminate the NL field $\mathscr{B}$ by integrating out it in the beginning, since the composite operator $\mathcal{O}$ defined by (\ref{op})  does not depend on $\mathscr{B}$.  Even after the elimination of $\mathscr{B}$, the total action is invariant under another BRST transformation which we call the {\it on-shell} BRST transformation.  The explicit form of the on-shell BRST transformation depends on the choice of the gauge fixing condition.  
In the  generalized Lorentz gauge \cite{KMSI02}, the on-shell BRST transformation reads 
\begin{subequations}
\begin{align}
 \bm{\delta}_{\rm B} \mathscr{A}_\mu(x)
   & =\mathscr{D}_\mu[\mathscr{A}]\mathscr{C}(x) ,
 \\
 \bm{\delta}_{\rm B} \mathscr{C}(x)
   & =-{1 \over 2}g(\mathscr{C}(x) \times \mathscr{C}(x)) , \\
 \bm{\delta}_{\rm B} \bar{\mathscr{C}}(x) 
 &= - \left[   {1 \over \lambda} \partial_\mu \mathscr{A}_\mu(x) + \xi  g \mathscr{C}(x) \times \bar{\mathscr{C}}(x)  \right] ,
\end{align}
\label{BRST2}
\end{subequations}
where $\lambda$ and $\xi$ are two gauge fixing parameters in the generalized Lorentz gauge. 
In the $\xi=0$, the generalized Lorentz gauge reduces to the ordinary Lorentz gauge as found in the textbook of quantum field theory where  a gauge fixing parameter $\lambda=0$ corresponds to the Landau gauge.  
It has been shown \cite{Kondo01} that the composite operator $\mathcal{O}$ defined by (\ref{op}) is on-shell BRST invariant, if the theory is restricted to the subspace specified by $\lambda=0$ or $\xi=1/2$.  
It is confirmed that this property is preserved under the renormalization \cite{KMSI02} and that the renormalization group flow is restricted to the subspace if it starts at a point in the subspace (In particular, $\lambda=0$ is a infrared fixed point).  
In what follows, we restrict our analysis to the first case of $\lambda \rightarrow 0$ with $\xi=0$ for simplicity.  
The second case will be discussed elsewhere.

\subsection{Step 2: Non-Abelian Stokes theorem}

\begin{figure}[htbp]
\begin{center}
\[
\begin{array}{cl}
\begin{array}{c}
\includegraphics{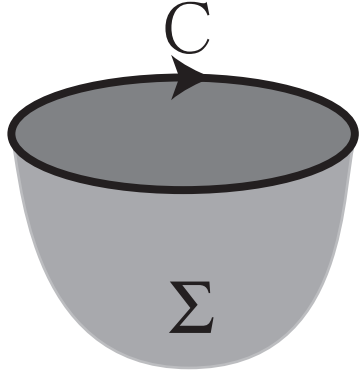}
\end{array}
\end{array}
\]
\caption{Wilson loop $C$ and the surface $\Sigma$.}
\label{fig:Wilsonloop1}
\end{center}
\end{figure}

First of all, we attempt to rewrite the non-Abelian Wilson loop operator into a quantity written in terms of a surface integral over an arbitrary surface $\Sigma$ with a boundary $C$, i.e., $\partial \Sigma =C$ (See Fig.~\ref{fig:Wilsonloop1}), by making use of a version \cite{DP89} of the non-Abelian Stokes theorem (NAST) \cite{NAST}.
In this paper, we adopt the Diakonov-Petrov (DP) version \cite{DP89} of NAST which does not contain the ordering of the surface integral,
although there are a number of versions of NAST, see references cited in \cite{KT99}.  
 The DP version of the NAST was first obtained for $G=SU(2)$ and the same result was rederived for $G=SU(2)$ by making use of the coherent state representation \cite{KondoIV} and later extended to $SU(N)$ by making use of the generalized coherent state representation \cite{KT99,HU99}.  
 For the simplest case of $G=SU(2)$, the NAST is given by  
\begin{equation}
  W_C[\mathscr{A}] = \int d\mu_S(V) \exp \left[ ig{J \over 2} \int_{\Sigma} dS^{\mu\nu} f_{\mu\nu}^{V} \right] , 
\end{equation}
where a character $J$ specifies the  $(2J+1)$-dimensional
    representation to which the probe source of quark for defining the Wilson loop operator belongs ($J=1/2$ for the fundamental representation, $J=1$ for the adjoint representation, and so on). 
Here $d\mu_S(V)$ is the product measure $\prod_{x\in S} d\mu(V(x))$ of the invariant Haar measure $d\mu(V(x))$ on the coset $G/H=SU(2)/U(1) (\cong S^2 \ni V)$ which is defined at an arbitrary point $x$ on the surface $S$.
For an infinitesimal surface element $dS^{\mu\nu}$, an antisymmetric tensor field $f_{\mu\nu}^{V}$ is defined by%
\footnote{
This tensor has the same form as the 't Hooft tensor for describing the 't Hooft-Polyakov magnetic monopole, if $V^A(x)$ is identified with a unit isovector of the Higgs scalar field, $\hat{\phi}^A := \phi^A/|\phi|$.
This fact suggests a possible interplay between magnetic monopole and quark confinement.  In other words, the magnetic monopole might be encoded in the Wilson loop operator, see \cite{KondoII,KondoIV,KondoV,KondoVI}.
}
\begin{align}
  f_{\mu\nu}^{V}(x) := \partial_\mu[{\bf V}(x) \cdot \mathscr{A}_\nu(x)] - \partial_\nu[{\bf V}(x) \cdot \mathscr{A}_\mu(x)]
  - g^{-1} {\bf V}(x) \cdot (\partial_\mu{\bf V}(x) \times \partial_\nu{\bf V}(x)) ,
\label{f}
\end{align}
where ${\bf V}$ is the flag variable ${\bf V}(x):=V^A(x) \sigma^A/2$ with Pauli matrices $\sigma^A (A=1,2,3)$ and three components of $V^A (A=1,2,3)$ constitute a unit vector, i.e., 
${\bf V}(x) \cdot {\bf V}(x) = V^A(x) V^A(x) = 1$.  Here the last term in (\ref{f}) corresponds to the topological term. 
By parameterizing a unit vector ${\bf V}$ explicitly as  
$V_1=\sin\theta \cos\varphi, V_2=\sin\theta \sin\varphi, V_3=\cos\theta$, the invariant Haar measure on $SU(2)/U(1)$ reads
\begin{equation}
  d\mu(V(x)) = \delta({\bf V}(x)\cdot{\bf V}(x)-1) dV_1(x) dV_2(x) dV_3(x)
  = \sin \theta(x) d\theta(x) d\varphi(x) .
\end{equation}
The $f_{\mu\nu}^{V}$ is invariant under the non-Abelian full gauge transformation, i.e.,
$\delta_{\omega} f_{\mu\nu}^{V}(x)=0$
for $\delta_{\omega} \mathscr{A}_{\mu}(x)=\mathcal{D}_\mu \omega(x)$
and $\delta_{\omega}{\bf V}(x)=ig[\omega(x),{\bf V}(x)]$ in the infinitesimal form. It is invariant also for a finite gauge transformation:
$\mathscr{A}_\mu(x) \rightarrow U(x) \mathscr{A}_\mu(x) U^\dagger(x)+ig^{-1} U(x) \partial_\mu U^\dagger(x)$, and 
${\bf V}(x) \rightarrow U(x){\bf V}(x)U^\dagger(x)$ for 
$U(x)=e^{ig\omega(x)}$ with $\omega(x)=\omega^A(x) \sigma^A/2$.
Similarly, the BRST transformations of ${\bf V}$ is defined by replacing $\omega$ with the ghost field $\mathscr{C}$, i.e., 
\begin{equation}
 \bm{\delta}_{\rm B}{\bf V}(x)=ig[ \mathscr{C}(x),{\bf V}(x)]
= g  {\bf V}(x) \times \mathscr{C}(x) ,
\end{equation}
just as the BRST transformation of $\mathscr{A}_\mu$ is given by
$\bm{\delta}_{\rm B}\mathscr{A}_{\mu}(x)=\mathscr{D}_\mu \mathscr{C}(x) .
$
Then the Wilson loop average reads
\begin{align}
  W(C)  
  = Z_{YM}^{-1} \int d\mu_{YM} e^{-S_{YM}^{tot}} \int d\mu_S(V) \exp \left[ ig{J \over 2} \int_{\Sigma} dS^{\mu\nu} f_{\mu\nu}^{V} \right]
   .
\end{align}

\par
\subsection{Step 3: Introducing auxiliary antisymmetric tensor field}
Next, we introduce an antisymmetric tensor field%
\footnote{The antisymmetric tensor $h_{\mu\nu}$ has no index for internal symmetry.  
It is possible to introduce the antisymmetric tensor field $h_{\mu\nu}^A$ in the adjoint representation of the gauge group $G$.
However, we do not use this type of tensor.  The reason is as follows.  
The tensor $h_{\mu\nu}^A$ is not gauge invariant.  Therefore, the Abelian components are identified with the diagonal components and selected out by making use of a naive procedure of Abelian projection, see Ellwanger \cite{Ellwanger98}.  
However, this procedure breaks the global color symmetry.  
}
 $h_{\mu\nu}(x)$ (of canonical mass dimension 2) by inserting an identity into the functional integral: 
\begin{equation}
  1 = \int \mathcal{D}h_{\mu\nu} \exp \left\{ -{1 \over 4\kappa_0} \int d^4x [h_{\mu\nu}(x)-f^V_{\mu\nu}(x)]^2 \right\} ,
  \label{insert}
\end{equation}
where $\kappa_0$ is an arbitrary dimensionless constant and the normalization factor is included in the measure%
\footnote{
For the integration by parts to be possible, the translational invariance of the measure is assumed to hold, i.e.,
$\mathcal{D}(h+a)=\mathcal{D}h$, which implies an identity, 
$
 \int \mathcal{D}h {\delta \over \delta h_{\mu\nu}}f(h)=0  
$
for arbitrary functional $f(h)$ of $h$.
This is in fact necessary to define the Gaussian integration (\ref{insert}).
The constant $\kappa_0$ will be determined so as to reproduce the correct normalization of the dual Ginzburg-Landau theory in the final stage of the derivation.
}
$\mathcal{D}h_{\mu\nu}$.
The simplest way to keep the gauge and/or BRST invariances 
is to require the relation, 
$\delta_\omega [h_{\mu\nu}(x)- f_{\mu\nu}^{V}(x)]=0$
and/or
$\bm{\delta}_{\rm B} [h_{\mu\nu}(x)- f_{\mu\nu}^{V}(x)]=0$.
Hence, the BRST invariance is preserved, if $h_{\mu\nu}$ is an Abelian tensor invariant under the BRST transformation, i.e.,
\begin{equation}
\bm{\delta}_{\rm B} h_{\mu\nu}(x)=0 ,
\end{equation}
since
$\bm{\delta}_{\rm B} f_{\mu\nu}^{V}(x)=0$.
\par
It should be remarked that the tensor field $h_{\mu\nu}$ is an auxiliary field without its kinetic term at this state.
However, the kinetic term will be generated by radiative corrections in the quantum theory.  This is one of the most important points in deriving the string representation of the Yang-Mills theory, as will be discussed in the following.  
\par
For (\ref{insert}) to be meaningful,  furthermore, ${\bf V}(x)$ must be defined on the whole space ${\bf R}^4$.  Therefore, the field ${\bf V}(x)$ on the surface $\Sigma$ must be continued to the whole space ${\bf R}^4$ outside the surface $\Sigma$.  This continuation would not cause a trouble, since this continuation does not lead to the failure of uniqueness due to encountering the singularity (Such a possibility was carefully examined when the line integral is continued into the surface integral in NAST by Diakonov and Petrov \cite{DP01}).%
\footnote{
The surface integral is obtained from the line integral of the diagonal component $a_\mu :={\rm tr}[\sigma_3 \mathscr{A}_\mu^U]$ of 
$\mathscr{A}_\mu^U:=U \mathscr{A}_\mu U^\dagger+g^{-1}iU \partial_\mu U^\dagger$ by the usual Stokes theorem as
$
 \oint_{C} dx^\mu a_\mu  = \int_{\Sigma} dS^{\mu\nu} f_{\mu\nu}^V .
$
Hence $V^A(x) = {\rm tr}[U^\dagger(x) \sigma_3 U(x)\sigma^A ]$ for
$
 U(x)=\exp (i\chi(x)\sigma_3/2)\exp (i\theta(x)\sigma_2/2)\exp (i\varphi(x)\sigma_3/2) \in SU(2) ,
$
where $\chi,\theta,\varphi$ are Euler angles.
Note that $V^A$ is invariant under the residual U(1) gauge transformation and hence it is the flag variable \cite{KT99} belonging to $SU(2)/U(1)$.
See \cite{KondoIV,KT99} for details.
}
  Therefore we can replace $d\mu_S(V)$ with $d\mu_{R^4}(V):=d\mu_{\infty}(V)$ without any difficulty 
(This issue will be further discussed in a forthcoming paper).
\par
Thus the Wilson loop average is written as
\begin{align}
  W(C)  
  =& \int \mathcal{D}h_{\mu\nu} 
  \exp \left\{ -{1 \over 4\kappa_0} \int d^4x  h_{\mu\nu}^2(x) \right\}
  Z_{YM}^{-1} \int d\mu_{YM} e^{-S_{YM}^{tot}} 
  \nonumber\\
  & \times \Big\langle
  \exp \left\{  {1 \over \kappa_0} \int d^4x \left( {1 \over 2}h_{\mu\nu} f^V_{\mu\nu}-{1 \over 4}f^V_{\mu\nu}f^V_{\mu\nu} \right) \right\}
  \exp \left[ ig{J \over 2} \int_{S} dS^{\mu\nu} f_{\mu\nu}^{V} \right] \Big\rangle_V
   ,
\end{align}
where we have introduced the expectation functional, 
\begin{equation}
\langle [\cdots] \rangle_V:=\int d\mu_{\infty}(V) [\cdots] .
\end{equation}

\par
\subsection{Step 4: Cumulant expansions}
By introducing the {\it vorticity tensor current} $\Theta_{\mu\nu}(x)$ by
\begin{equation}
  \Theta_{\mu\nu}(x) := \int_{\Sigma} d^2S_{\mu\nu}(x(\sigma)) \delta^4(x-x(\sigma)) ,
\end{equation}
it is possible to cast the surface integral into the space-time integral according to 
$
  \int_{\Sigma} dS^{\mu\nu}(x(\sigma)) f_{\mu\nu}^{V}(x(\sigma))
  = \int d^4x \Theta_{\mu\nu}(x) f_{\mu\nu}^{V}(x) ,
$
where $\Theta_{\mu\nu}(x)$ has its support on the two-dimensional surface $\Sigma$ (parameterized by the coordinate $\sigma=(\sigma_1,\sigma_2)$) with a target space coordinate $x_\mu=x_\mu(\sigma)$.  
We attempt to convert the interaction term of the surface element $d\sigma^{\mu\nu}$ with the $f_{\mu\nu}^{V}$ into that of the surface element $d\sigma^{\mu\nu}$ with $h_{\mu\nu}$.  
In fact,  an identity:
\begin{align}
&   \exp \left[ ig{J \over 2} \int_{S} d^2S^{\mu\nu} f_{\mu\nu}^{V} \right]  
  \exp \left[  {1 \over \kappa_0} \int d^4x  {1 \over 2}h_{\mu\nu} f^V_{\mu\nu} \right]
   \nonumber\\ =&
   \exp \left[ ig{J \over 2} \int d^4x \Theta_{\mu\nu} f_{\mu\nu}^{V} \right] \exp \left[  {1 \over \kappa_0} \int d^4x {1 \over 2}h_{\mu\nu} f_{\mu\nu}^{V} \right]  
   \nonumber\\ =&
   \exp \left[ igJ \kappa_0 \int d^4x \Theta_{\mu\nu}{\delta \over \delta h_{\mu\nu}} \right] \exp \left[  {1 \over \kappa_0} \int d^4x {1 \over 2}h_{\mu\nu} f_{\mu\nu}^{V} \right] ,
\end{align}
and the integration by parts allows us to achieve the desired result: 
\begin{align}
  W(C)  
  =& \int \mathcal{D}h_{\mu\nu} \left( \exp \left[ igJ \kappa_0 \int d^4x \Theta_{\mu\nu}{\delta \over \delta h_{\mu\nu}} \right] 
  \exp \left\{ -{1 \over 4\kappa_0} \int d^4x  h_{\mu\nu}^2 \right\} \right)
  \nonumber\\
  & \times   Z_{YM}^{-1} \int d\mu_{YM} e^{-S_{YM}^{tot}} 
\Big\langle
  \exp \left\{  {1 \over \kappa_0} \int d^4x \left( {1 \over 2}h_{\mu\nu} f^V_{\mu\nu}-{1 \over 4}f^V_{\mu\nu}f^V_{\mu\nu} \right) \right\}
   \Big\rangle_V .
\end{align}
\par

It should be remarked that the ${\bf V}$ field is an important but artificial field (without any direct physical relevance) which has been introduced in the NAST to rewrite the line integral into the surface integral.  
Now we attempt to eliminate (or integrate out) the ${\bf V}$ field 
to obtain an effective theory written in terms of the auxiliary field $h_{\mu\nu}$ and the fundamental field 
$\Phi=\{ \mathscr{A}_\mu, \mathscr{B}, \mathscr{C}, \mathscr{\bar{C}} 
\}$.  
In the formal level, this step can be realized as
\begin{align}
  W(C)  
  =&  Z_{YM}^{-1} \int d\mu_{YM} \int \mathcal{D}h_{\mu\nu} 
  \exp \left[ igJ \kappa_0 \int d^4x \Theta_{\mu\nu}{\delta \over \delta h'{}_{\mu\nu}} \right]
  \nonumber\\
 & \times  \exp \left\{ -{1 \over 4\kappa_0} \int d^4x  h'{}_{\mu\nu}^2 
  -S_{YM}^{tot}[\Phi] + F[h,\mathscr{A}] \right\} \Big|_{h'=h} ,
\end{align}
where we have introduced $F[h,\mathscr{A}]$ which is written as the formal power series in $\kappa_0^{-1}$ by making use of the cumulant expansion for the expectation value $\langle (\cdots) \rangle_V$:
\begin{align}
  F[h,\mathscr{A}] 
=& \ln \Big\langle
  \exp \left\{  {1 \over \kappa_0} \int d^4x \left( {1 \over 2}h_{\mu\nu} f^V_{\mu\nu}-{1 \over 4}f^V_{\mu\nu}f^V_{\mu\nu} \right) \right\}
   \Big\rangle_V 
\nonumber\\
=&   \sum_{n=1}^{\infty} {\kappa_0^{-n} \over n!} 
  \Big\langle \left[  \int d^4x  {1 \over 2}h_{\mu\nu} f^V_{\mu\nu} 
  - \int d^4x {1 \over 4}f^V_{\mu\nu}f^V_{\mu\nu}   
  \right]^n \Big\rangle^{\text{con}}_V ,
\end{align}
with the connected expectation value $\langle (\cdots) \rangle_V^{\text{con}}$  defined from $\langle (\cdots) \rangle_V$ in the usual way.

\par
Subsequently, integrating out all the fundamental fields $\Phi$ leaving $h$ untouched, we 
arrive at the effective theory written in terms of an antisymmetric tensor field $h_{\mu\nu}$ alone.  
In addition to the first cumulant expansion for $\langle (\cdots) \rangle_V$:
\begin{align}
  W(C)  
  =&  \int \mathcal{D}h_{\mu\nu} 
  \left( \exp \left[ igJ \kappa_0 \int d^4x \Theta_{\mu\nu}{\delta \over \delta h_{\mu\nu}} \right] 
  e^{ -{1 \over 4\kappa_0} \int d^4x  h_{\mu\nu}^2 } \right)
    \Big\langle  
  e^{  F[h,\mathscr{A}] }
   \Big\rangle_{YM} ,
\end{align}
the second cumulant expansion for $\langle (\cdots) \rangle_{YM}$ results in 
\begin{align}
  W(C) =&  \int \mathcal{D}h_{\mu\nu} 
    \exp \left[ igJ \kappa_0 \int d^4x \Theta_{\mu\nu}{\delta \over \delta h'{}_{\mu\nu}} \right]
 e^{-S_h } \Big|_{h'=h} ,
  \\
  S_h :=&  {1 \over 4\kappa_0} \int d^4x  h'{}_{\mu\nu}^2  
 - \sum_{m=1}^{\infty} {1 \over m!} 
    \Big\langle  
    \left(  F[h,\mathscr{A}] \right)^m  
   \Big\rangle_{YM}^{\text{con}} ,
\end{align}
where $\langle (\cdots) \rangle_{YM}^{\text{con}}$ denotes the connected expectation value in the Yang-Mills theory.  

\par
For the purpose of obtaining the effective theory written in terms of $h$ alone without breaking the BRST invariance, it is better to use the form:
\begin{align}
  W(C)  
  =& \int \mathcal{D}h_{\mu\nu} \left( \exp \left[ igJ \kappa_0 \int d^4x \Theta_{\mu\nu}{\delta \over \delta h_{\mu\nu}} \right] 
  \exp \left\{ -{1 \over 4\kappa_0} \int d^4x  h_{\mu\nu}^2 \right\} \right)
  \nonumber\\
  &  \times
\Big\langle
\Big\langle
  \exp \left\{  {1 \over \kappa_0} \int d^4x \left( {1 \over 2}h_{\mu\nu} f^V_{\mu\nu}-{1 \over 4}f^V_{\mu\nu}f^V_{\mu\nu} \right) \right\}
   \Big\rangle_V 
\Big\rangle_{YM} .
\end{align}
Therefore the cumulant expansion reads
\begin{align}
  W(C)  
  = \int \mathcal{D}h_{\mu\nu} 
  \exp \left[ igJ \kappa_0 \int d^4x \Theta_{\mu\nu}{\delta \over \delta h'{}_{\mu\nu}} \right]
  \exp \left\{ -{1 \over 4\kappa_0} \int d^4x  h'{}_{\mu\nu}^2  + \mathcal{G}[h] \right\} \Big|_{h'=h} ,
\end{align}
by making use of the cumulant expansion for the double expectation $\langle \langle (\cdots) \rangle_V \rangle_{YM}$ defined by the formal power series in $\kappa_0^{-1}$:
\begin{align}
 \mathcal{G}[h]
=& \ln \Big\langle
\Big\langle
  \exp \left\{  {1 \over \kappa_0} \int d^4x \left( {1 \over 2}h_{\mu\nu} f^V_{\mu\nu}-{1 \over 4}f^V_{\mu\nu}f^V_{\mu\nu} \right) \right\}
   \Big\rangle_V 
\Big\rangle_{YM} 
\nonumber\\ 
=&   \sum_{n=1}^{\infty} {\kappa_0^{-n} \over n!} 
  \Big\langle
 \Big\langle \left[  \int d^4x  {1 \over 2}h_{\mu\nu} f^V_{\mu\nu} 
  - \int d^4x {1 \over 4}f^V_{\mu\nu}f^V_{\mu\nu}   
  \right]^n \Big\rangle^{\text{con}}_V \Big\rangle^{\text{con}}_{YM}  .
  \label{dcex}
\end{align}
\par
Furthermore, it is possible to calculate the derivative with respect to $h$ in the closed form,%
\footnote{We have used the Baker-Campbell-Hausdorff formula
\begin{equation}
  e^X e^Y = \exp \left\{ X+Y+{1 \over 2}[X,Y]+{1 \over 12}([X,[X,Y]]+[Y,[Y,X]]) + \cdots  \right\} .
\end{equation}
}
\begin{align}
 & \exp \left[ igJ \kappa_0 \int d^4y \Theta_{\mu\nu}(y) {\delta \over \delta h_{\mu\nu}(y)} \right] 
  e^{ -{1 \over 4\kappa_0} \int d^4x  h_{\mu\nu}^2(x) }
\nonumber\\
=& \exp \left[ 
-{1 \over 4\kappa_0} \int d^4x  h_{\mu\nu}^2(x) 
-{i \over 4}gJ \int d^4x \Theta_{\mu\nu}(x) h_{\mu\nu}(x)
+ {1 \over 24} (gJ)^2 \kappa_0 \int d^4x \Theta_{\mu\nu}^2(x) 
 \right] .
\end{align}
This procedure is not necessarily an indispensable step, but it is useful to avoid the complicated calculations in the later stage of integrating out all the fields except for $h$.
Thus we obtain
\begin{align}
  W(C) =&  \int \mathcal{D}h_{\mu\nu}  
e^{-S_h + {\kappa_0 \over 24} (gJ)^2  \int d^4x \Theta_{\mu\nu}^2(x) } 
    \exp \left[ -ig{J \over 2}  \int d^4x \Theta_{\mu\nu} h_{\mu\nu} 
\right] ,
  \\
  S_h :=&  {1 \over 4\kappa_0} \int d^4x  h_{\mu\nu}^2  
 - \mathcal{G}[h]  .
\end{align}
In this step, we are interested in only the terms coupled to the $h_{\mu\nu}$ field, since other terms merely give $h_{\mu\nu}$-independent constants after taking the double expectations or double cumulants, 
$\langle (\cdots) \rangle_V^{\text{con}}$ and $\langle (\cdots) \rangle_{YM}^{\text{con}}$.  
The full form of the action written in terms of $\Phi$ and $h$ is quite complicated but necessary for quantitative analysis and will be given in a subsequent paper.

\par
Retaining only the terms up to quadratic in $h$  in the cumulant expansion (which we call the {\it weak field approximation} \cite{Polyakov86} or the {\it bilocal approximation}),%
\footnote{It turns out that the quadratic terms are sufficient to reproduce the London limit in the type II superconductor.  However, we can show that the quartic term in $h$ is necessary to go beyond the London limit and to reach the border between type II and type I from the type II, see Appendix \ref{sec:dual}.
Especially, the truncation of the series excludes the possibility for the series in $h$ to be resumed into a closed form which might be a periodic function of $h$. If so, it will be possible to distinguish the half-integer $J$  representation from the integer $J$ one in the calculation of the string tension, as realized in compact QED \cite{Polyakov86}.
}
we obtain an expression of the effective theory written in terms of the $h$ alone (see Appendix \ref{sec:cumulant} for details of calculations):
\begin{align}
  S_h =& \int d^4x  {1 \over 4\kappa_0} h_{\mu\nu}(x)h_{\mu\nu}(x) 
-  {1 \over 8\kappa_0^2}\int d^4x \int d^4y 
h_{\mu\nu}(x) h_{\rho\sigma}(y)  g^{-2} 
\langle \Omega_{\mu\nu}(x)  \Omega_{\rho\sigma}(y) \rangle_{V}^{\text{con}} 
  \nonumber\\&
  -  {1 \over 2\kappa_0^2}  \int d^4x \int d^4 y
 \langle V^A(x) V^B(y) \rangle_{V}^{\text{con}} \langle \mathscr{A}_\mu^A(x) \mathscr{A}_\nu^B(y) \rangle_{YM}^{\text{con}} 
 \partial_\alpha h_{\mu\alpha}(x) \partial_\beta h_{\nu\beta}(y)
 + \cdots  ,
 \label{nlaction}
\end{align}
where we have introduced an antisymmetric tensor of canonical mass dimension 2:
\begin{equation}
\Omega_{\mu\nu}(x) :={\bf V}(x) \cdot (\partial_\mu{\bf V}(x) \times \partial_\nu{\bf V}(x)) = - \Omega_{\nu\mu}(x).
\end{equation}

\par
\subsection{Step 5: Gluon condensation and dynamical generation of the kinetic term for the antisymmetric tensor field}

\par
 First, we assume the translational invariance for the VEV 
$\langle (\cdots) \rangle_{V}$.  Hence we have  
\begin{equation}
 \langle V^A(x) V^B(y) \rangle_{V}^{\text{con}} =  G^{AB}(x-y) 
\label{G}  
\end{equation}
with a normalization $G^{AA}(0)=1$. 
In addition, we can put
\begin{equation}
 \langle \Omega_{\mu\nu}(x)  \Omega_{\rho\sigma}(y) \rangle_{V}^{\text{con}} 
   = K_{\mu\nu\rho\sigma}(x-y) ,
\label{K}
\end{equation}
where $K_{\mu\nu\rho\sigma}$ has the symmetry:
$
 K_{\mu\nu\rho\sigma} = - K_{\nu\mu\rho\sigma} = - K_{\mu\nu\sigma\rho} = K_{\rho\sigma\mu\nu} .
$
 Note that 
$
 \langle V^A(x) \rangle_{V}^{\text{con}} =  0 ,  
$
and
$
 \langle \Omega_{\mu\nu}(x)  \rangle_{V}^{\text{con}} = 0 .
$ 

Second, we expand $h(x)$ and $h(y)$ in powers of the relative coordinate $r_\mu := (x-y)_\mu$ around the center-of-mass coordinate $X:=(x+y)/2$.  Since  
 $x=X+r/2, y=X-r/2$,
 we have
\begin{equation}
 \begin{Bmatrix} h_{\mu\nu}(x) \\
                 h_{\mu\nu}(y) 
 \end{Bmatrix}
= h_{\mu\nu}(X) \pm {1 \over 2}r_\alpha \partial_\alpha h_{\mu\nu}(X) + {1 \over 8} r_\alpha r_\beta \partial_\alpha \partial_\beta h_{\mu\nu}(X) + \cdots .
\end{equation}

Third, we take into account the decomposition of the double integration over the space-time:
\begin{equation}
\int d^4x \int d^4y (\cdots) = \int d^4 r \int d^4X (\cdots) ,
\end{equation}
where the Jacobian for change of variables is equal to one.

\par
The first term of $S_h$ (\ref{nlaction}) is the mass term for $h_{\mu\nu}$.
The second term of $S_h$ (\ref{nlaction}) is cast into 
\begin{align}
  & \int d^4x \int d^4y 
h_{\mu\nu}(x) h_{\rho\sigma}(y) 
\langle \Omega_{\mu\nu}(x)  \Omega_{\rho\sigma}(y) \rangle_{V}^{\text{con}}   \nonumber\\ 
  =& \int d^4r K_{\mu\nu\rho\sigma}(r) 
\int d^4X  h_{\mu\nu}(X) h_{\rho\sigma}(X)
\nonumber\\
&+ \int d^4r K_{\mu\nu\rho\sigma}(r) {1 \over 2} r_\alpha r_\beta 
\int d^4X  h_{\mu\nu}(X) \partial_\alpha \partial_\beta h_{\rho\sigma}(X)  + O(h \partial^4 h) 
\nonumber\\
  =& g^2  \omega^{(0)}
\int d^4X  h_{\mu\nu}(X) I_{\mu\nu\rho\sigma} h_{\rho\sigma}(X)
\nonumber\\
&+   {1 \over 8} g^2 \omega^{(1)}
\int d^4X  h_{\mu\nu}(X) I_{\mu\nu\rho\sigma} \partial^2 h_{\rho\sigma}(X) 
+ O(h \partial^4 h) ,
\label{2ndS}
\end{align}
where we have put 
\begin{equation}
  \int d^4r (r^2)^{\ell} K_{\mu\nu\rho\sigma}(r) 
= g^2 \omega^{(\ell)} I_{\mu\nu\rho\sigma}  ,
\quad
I_{\mu\nu\rho\sigma}:={1 \over 2}(\delta_{\mu\rho} \delta_{\nu\sigma} - \delta_{\mu\sigma} \delta_{\nu\rho}) .
\end{equation}
In the {\it tree level} (without loop corrections), the kinetic term for the field $V^A(x)$ 
is absent.%
\footnote{
In order to make the expectation value 
$\langle (\cdots) \rangle_{V}$
well-defined, however, we need the kinetic term of the $V$ field which comes out in the course of the derivation of NAST and goes to zero in the limit of removing the regularization, see \cite{DP89,KondoIV} for details.
Such a kinetic term can also be generated by radiative corrections,
as will be examined in a subsequent paper. 
}
 Hence $V^A(x)$ is subject to the short-range correlation,%
\footnote{The support of the relevant correlation function consists of a point $x=y$, if the kinetic term is absent. Therefore, it must be proportional to the delta function (and its derivatives).    
This is because for $x\not=y$, the absence of the kinetic term implies
$
 \langle V^A(x) V^B(y) \rangle_{V}^{\text{con}}
= \langle V^A(x) \rangle_{V}^{\text{con}} \langle  V^B(y) \rangle_{V}^{\text{con}} = 0 ,
$
and 
$
 \langle \Omega_{\mu\nu}(x)  \Omega_{\rho\sigma}(y) \rangle_{V}^{\text{con}} 
= \langle \Omega_{\mu\nu}(x) \rangle_{V}^{\text{con}} \langle  \Omega_{\rho\sigma}(y) \rangle_{V}^{\text{con}}
= 0 .
$
} 
$
 \langle V^A(x) V^B(y) \rangle_{V}^{\text{con}}  = C \delta^{AB} \delta^4(r) ,
$
where $C$ must be a constant with mass dimension $-4$ based on dimensional analysis, i.e., $C =C_0 \Lambda_{IR}^{-4}$ with a dimensionless constant $C_0$ (In other words, $\Lambda_{IR}^{-4}=\int d^4x$ is the volume of space-time).
The correlation of $\Omega_{\mu\nu}$ is also ultra-local:
$
 \langle \Omega_{\mu\nu}(x)  \Omega_{\rho\sigma}(y) \rangle_{V}^{\text{con}} 
   = C' I_{\mu\nu\rho\sigma}\delta^4(r) ,
$
where $C'$ is another dimensionless constant.
Therefore, we have $\omega^{(0)}\not=0$ and $\omega^{(\ell)}=0$ for $\ell \ge 1$ in the tree level.  This implies that the kinetic term for $h$ is not generated from the second term in the RHS of (\ref{nlaction}) in the tree level. 
 The contribution from $\omega^{(0)}\not=0$  renormalizes the mass term and can be absorbed by the redefinition of $\kappa_0$.
\par
It is possible to relate the action (\ref{nlaction}) to the conventional gluon condensate of mass dimension 4 \cite{SVZ79}, as will be discussed in the next section.  
On the contrary, we discuss in this section how to relate the action (\ref{nlaction}) to a novel vacuum condensate of mass dimension 2 \cite{Boucaudetal00,GSZ01,Kondo01}.
A pair of gluon field operators has the operator product expansion (OPE):%
\footnote{
It should be remarked that the gluon condensate of mass dimension 4, i.e., $\langle \mathcal{F}_{\mu\nu}^2(x) \rangle$ does not appear in the OPE of the gluon propagator, see \cite{LO92}.  Therefore, we cannot use it as a basic ingredient to derive the mass gap and quark confinement in our strategy.
}
\begin{align}
  \mathscr{A}_\mu^A(x) \mathscr{A}_\nu^B(y)
  =&   D_{\mu\nu}^{AB}(r) {\bf 1} + W^{[{1 \over 2}\mathscr{A}\mathscr{A}]}{}_{\mu\nu}^{AB}(r) \left[{1 \over 2}\mathscr{A}_\rho \cdot \mathscr{A}_\rho\left(X \right) \right] 
\nonumber\\&
 +  W^{[\bar{\mathscr{C}}\mathscr{C}]}{}_{\mu\nu}^{AB}(r) 
  \left[   \bar{\mathscr{C}} \cdot \mathscr{C}\left(X \right) \right] + \cdots ,
\label{gluonOPE}
\end{align}
where the first term $D_{\mu\nu}^{AB}$ is the gluon propagator dressed by the perturbative quantum corrections with the tree expression given by 
$
  (D_0)_{\mu\nu}^{AB}(r) = \delta^{AB}(1/\partial^2) [ \delta_{\mu\nu} -(1-\lambda) \partial_\mu \partial_\nu/\partial^2 ] \delta^4(r) .
$
\par
In this section, we focus on the Lorentz gauge condition $\partial_\mu \mathscr{A}_\mu(x)=0$ as a manifestly Lorentz covariant gauge fixing.
(Other types of gauge fixing will be examined in the forthcoming paper.)
In the Landau gauge $\lambda=0$,  the  Wilson coefficient has been calculated in \cite{Boucaudetal00,Boucaudetal01}:  
$W^{[{1 \over 2}\mathscr{A}\mathscr{A}]}{}_{\mu\nu}^{AB} \not=0$, while 
$W^{[i\bar{\mathscr{C}}\mathscr{C}]}{}_{\mu\nu}^{AB}=0$.%
\footnote{
In gauges other than the Landau, the operator mixing among the operators with the same mass dimensions and the same symmetries does take place so that $W^{[i\bar{\mathscr{C}}\mathscr{C}]}{}_{\mu\nu}^{AB}\not=0$, as demonstrated in \cite{KMSI02}.  
Only in the Landau gauge $\lambda=0$, the operator mixing disappears.  
The ghost-antighost composite operator mixes with the gluon pair composite operator under the renormalization.  
Here a BRST invariant combination of two composite operators 
appears in the calculation beyond the tree level.  See \cite{KMSI02} for details.
}
This result was also confirmed as a very special limit in the framework of the most general Lorentz gauge fixing \cite{Kondo01,KMSI02} where
$W^{[i\bar{\mathscr{C}}\mathscr{C}]}{}_{\mu\nu}^{AB}\not=0$ in general. 
 For $SU(N_c)$ Yang-Mills theory, the RG improved value of the Wilson coefficient is given by
$
 \tilde{W}^{[{1 \over 2}\mathscr{A}\mathscr{A}]}{}_{\mu\nu}^{AB}(p)
= \delta^{AB} \tilde T_1^{\mu\nu}(p)
({\ln p/\Lambda_0 \over \ln \mu/\Lambda_0})^{{ 3\over 4}{N_c \over \beta_0}}
$
 in momentum representation 
where $\beta_0:={11 \over 3}N_c$ and 
$\tilde T_1^{\mu\nu}$ is the tree result:
$\tilde T_1^{\mu\nu}(p)= {N_c g^2 \over 2(N_c^2-1)} [(\tilde D_0)^2P_T]_{\mu\nu}$
with the transverse projection
$P_{\mu\nu}^T:=\delta_{\mu\nu}-p_\mu p_\nu/p^2$.
Here $\Lambda_0$ is the RG invariant constant in Yang-Mills theory corresponding to $\Lambda_{QCD}$ in QCD.  
Therefore,
$
 \tilde{W}^{[{1 \over 2}\mathscr{A}\mathscr{A}]}{}_{\mu\nu}^{AB}(p)
= \delta^{AB} {N_c g^2 \over 2(N_c^2-1)}{1 \over p^4}
({\ln p/\Lambda_0 \over \ln \mu/\Lambda_0})^{{ 3\over 4}{N_c \over \beta_0}} P_{\mu\nu}^T .
$
On the other hand, the RG improved gluon propagator is given by
$
 \tilde D_{\mu\nu}^{AB}(p) = ({\ln p/\Lambda_0 \over \ln \mu/\Lambda_0})^{{13 \over 6}{N_c \over \beta_0}} (\tilde D_0)_{\mu\nu}^{AB}(p) ,
$
see \cite{Boucaudetal01,KMSI02}.
\par
Hence the third term of (\ref{nlaction}) reads 
\begin{align}
  & \int d^4x \int d^4 y
 \langle V^A(x) V^B(y) \rangle_{V}^{\text{con}} \langle \mathscr{A}_\mu^A(x) \mathscr{A}_\nu^B(y) \rangle_{YM}^{\text{con}} 
 \partial_\alpha h_{\mu\alpha}(x) \partial_\beta h_{\nu\beta}(y) ,
\nonumber\\
=&   \int d^4x \  \partial_\alpha h_{\mu\alpha}(x) {1 \over \partial^2} \partial_\beta h_{\mu\beta}(x)
\nonumber\\
&- {1 \over 4} \sum_{\ell=0}^{\infty} K^{(\ell)}_{\mu\nu}  
  \int d^4 X
  \Big\langle {1 \over 2}\mathscr{A}_\rho^2(X) \Big\rangle_{YM} 
 \partial_\alpha h_{\mu\alpha}(X) (\partial^2)^\ell \partial_\beta h_{\nu\beta}(X)   ,
\label{3rdterm}
\end{align}
where we have defined
\begin{equation}
 K_{\mu\nu}^{(\ell)} = a_\ell \int d^4r \ (r^2)^\ell 
  G^{AB}(r)  W^{[{1 \over 2}\mathscr{A}\mathscr{A}]}{}_{\mu\nu}^{AB}(r) ,
\label{defK}
\end{equation}
with appropriate numerical coefficients
$a_\ell (\ell=0, 1, 2, \cdots)$, in particular,
$a_0=1$. 
Note that the canonical mass dimension of $K_{\mu\nu}^{(\ell)}$ is $-(4+2\ell)$.
The first term of the right-hand side (RHS) of (\ref{3rdterm}) comes from the first term of RHS of (\ref{gluonOPE}) (without perturbative radiative corrections) and  contributes to the renormalization of the mass term, leading to  a modification of $\kappa_0$.
Hereafter the vacuum condensate 
$
  \langle {1 \over 2}\mathscr{A}_\rho^2(X) \rangle_{YM}
$
in (\ref{3rdterm}) 
should be understood to be an Euclidean quantity together with the Wilson coefficient 
$W^{[{1 \over 2}\mathscr{A}\mathscr{A}]}{}_{\mu\nu}^{AB}$ 
in (\ref{defK}). 

\par
The first two terms $\ell=0, 1$ in the series are sufficient for our purpose at moment.  
Note that $K_{\mu\nu}^{(0)}\not=0$ and $K_{\mu\nu}^{(\ell)}=0$ for $\ell \ge 1$  in the {\it tree} level.  
The term proportional to $\omega^{(0)}$ is absorbed into the redefinition of $\kappa_0$. 
Therefore, we have only to consider $K_{\mu\nu}^{(0)}$.
Thus we obtain the effective theory (up to higher power terms and higher derivative terms):
\begin{align}
  S_h =& \int d^4X  {1 \over 4\kappa_0} h_{\mu\nu}(X)h_{\mu\nu}(X) 
   - {1 \over 2\kappa_0^2} \int d^4X \  \partial_\alpha h_{\mu\alpha}(X) {1 \over \partial^2} \partial_\beta h_{\mu\beta}(X)
\nonumber\\
&  - \int d^4X {1 \over 64\kappa_0^2} \omega^{(1)}
 h_{\mu\nu}(X) I_{\mu\nu\rho\sigma} \partial^2 h_{\rho\sigma}(X)  
\nonumber\\
&   +  {1 \over 8\kappa_0^2} \sum_{\ell=0}^{1} K^{(\ell)}_{\mu\nu}  
  \int d^4 X
  \Big\langle {1 \over 2}\mathscr{A}_\rho^2(X) \Big\rangle_{YM} 
 \partial_\alpha h_{\mu\alpha}(X) (\partial^2)^\ell \partial_\beta h_{\nu\beta}(X) 
+ \cdots  .
\label{expan}
\end{align}
Even in the tree level, therefore, the kinetic term for $h$ is generated from the third term of (\ref{nlaction}) if the vacuum condensate of mass dimension 2 exists as a non-perturbative effect.%
\footnote{
On the other hand, if $V^A(x)$  was subject to the short-range correlation as in the tree level, 
the effective theory in $h$ was simply obtained without relying on the OPE as
\begin{equation}
  S_h = \int d^4x \left[ {1 \over 4\kappa} h_{\mu\nu}(x)h_{\mu\nu}(x) 
  +  {C \over \kappa_0^2}  {\delta_{\mu\nu} \over 4}
  \Big\langle {1 \over 2}\mathscr{A}_\rho^2(x) \Big\rangle_{YM} 
 \partial_\alpha h_{\mu\alpha}(x) \partial_\beta h_{\nu\beta}(x) 
 \right] ,
\end{equation}
where $C$ must be a constant with mass dimension $-4$ based on dimensional analysis, i.e., $C =C_0 \Lambda_{IR}^{-4}$ with a dimensionless constant $C_0$ and a dimensionful scale $\Lambda_{IR}$. 
This case is derived in (\ref{nlaction}) by taking the correlation:
$
 \langle V^A(x) V^B(y) \rangle_{V}^{\text{con}}  = C_0 \Lambda_{IR}^{-4} \delta^{AB} \delta^4(r) .
$
}
 Thus we have observed that {\it the kinetic term of $h_{\mu\nu}$ is generated by the existence of vacuum condensate of mass dimension 2}.
\par
This phenomena was already pointed out to occur in the Maximal Abelian (MA) gauge in the previous paper \cite{Kondo00} where the ghost--antighost condensation \cite{Schaden99,KS00} played the similar role to the gluon pair condensation.
In the previous paper, however, we have used from the beginning the massive propagator for the off-diagonal gluon in the MA gauge in the course of loop calculation.  In such a case, quadratic divergence may appear.
The derivation just given in this paper does not use such an assumption, although it is expected that the gluon pair condensation yields the massive gluon.  
Therefore, we can avoid the problem of quadratic divergence in our approach.  In fact, we can explicitly show that there is no quadratic divergence due to cancellation among Feynman diagrams (Fig.1, Fig.2) indicated in \cite{KMSI02}, provided that the BRST symmetry is not broken.

\par
\subsection{Step 6: Effective antisymmetric tensor theory}
In order to obtain a dual theory of the Yang-Mills theory, we introduce the dual variable $B_{\mu\nu}$ as the Hodge dual of
$h_{\mu\nu}$, i.e.,
\begin{equation}
 B_{\mu\nu} := \ {}^*h_{\mu\nu} \equiv {1 \over 2} \epsilon_{\mu\nu\rho\sigma} h^{\rho\sigma} .
\end{equation}
It is obvious that $B_{\mu\nu}$ is both gauge and BRST invariant.
Note that
\begin{align}
 h_{\mu\nu}h_{\mu\nu} =& B_{\mu\nu}B_{\mu\nu} ,
\\
 \delta_{\mu\nu} \partial_\alpha h_{\mu\alpha} \partial_\beta h_{\nu\beta} 
=& {1 \over 4} (\epsilon_{\mu\nu\rho\sigma}\partial^\nu B^{\rho\sigma})^2 .
\end{align}
Similarly, we have for any integer $\ell=-1, 0, 1, \cdots$
\begin{equation}
 \delta_{\mu\nu} \partial_\alpha h_{\mu\alpha} (\partial^2)^\ell \partial_\beta h_{\nu\beta} 
= 
{1 \over 4} (\epsilon_{\mu\nu\rho\sigma}\partial^\nu B^{\rho\sigma})
(\partial^2)^\ell (\epsilon_{\mu\alpha\beta\gamma}\partial^\alpha B^{\beta\gamma}) .
\end{equation}
Thus we arrive at a BRST-invariant dual theory which is equivalent to Yang-Mills theory with an insertion of the Wilson loop operator:
\begin{align}
  W(C) \cong& Z_{YM}^{-1} \int \mathcal{D}B_{\mu\nu} 
e^{-S_d[B] + {\kappa_0 \over 24} (gJ)^2  \int d^4x \Theta_{\mu\nu}^2(x) } 
    \exp \left[ -ig{J \over 2}  \int d^4x \ {}^*\Theta_{\mu\nu} B_{\mu\nu} \right] ,
\\
  S_d[B] =& \int d^4x  \Biggr[ {1 \over 4\kappa_0} B_{\mu\nu}^2 
  + {K^{(0)} \sigma_{vc} \over 32\kappa_0^2} (\epsilon^{\mu\nu\rho\sigma} \partial_\nu B_{\rho\sigma})^2 
  -  {1 \over 8\kappa_0^2} 
  (\epsilon_{\mu\nu\rho\sigma}\partial^\nu B^{\rho\sigma})
{1 \over \partial^2} (\epsilon_{\mu\alpha\beta\gamma}\partial^\alpha B^{\beta\gamma})
\nonumber\\&
  - {\omega^{(1)} \over 64\kappa_0^2} B_{\mu\nu} I_{\mu\nu\rho\sigma} \partial^2 B_{\rho\sigma} 
  - {K^{(1)} \sigma_{vc} \over 32\kappa_0^2} (\partial_\alpha \epsilon^{\mu\nu\rho\sigma} \partial_\nu B_{\rho\sigma})^2 
  + \cdots \Biggr]  ,
  \label{Bth}
\end{align}
where we have put
\footnote{
In the Landau gauge, $W^{[{1 \over 2}\mathscr{A}\mathscr{A}]}{}_{\mu\nu}$ is proportional to the transverse projection $P^T_{\mu\nu}:=\delta_{\mu\nu}-\partial_\mu \partial^{-2} \partial_\nu$, see \cite{Boucaudetal01,KMSI02}. 
Therefore, we can regard that the tensor structure of $K_{\mu\nu}^{(\ell)}$
is simply proportional to 
$\delta_{\mu\nu}$, since antisymmetry of $h_{\mu\nu}$ implies  
$\partial_\mu \partial_\alpha h_{\mu\alpha}=0$. 
} 
\begin{equation}
 K_{\mu\nu}^{(\ell)}:= K^{(\ell)} \delta_{\mu\nu} 
\end{equation}
by introducing dimensionless constants $K^{(\ell)}$ of canonical mass dimension $-(4+2\ell)$, 
and  the vacuum condensate $\sigma_{vc}$ of mass dimension 2 is defined by the VEV of a composite operator $\mathcal{O}$:
\begin{equation}
  \sigma_{vc} := \langle \mathcal{O} \rangle_{YM} .
\end{equation}
 For the conventional Lorentz gauge,   
the vacuum condensate has been evaluated only in the Landau gauge, see \cite{VKAV01} for the analytical result and  \cite{Boucaudetal01,Boucaudetal02} for the numerical results.  Both results conclude that the non-zero vacuum condensate $\sigma_{vc}$ does indeed exist where  
$
 \sigma_{vc} = \langle {1 \over 2}\mathscr{A}_\mu^2(0) \rangle
\not= 0  
$
for the translational invariant vacuum.
It is possible to calculate the vacuum condensate in the most general Lorentz gauge, see \cite{Kondo01,KMSI02}.  
\par

\par
\subsection{Step 7: Area law of the Wilson loop average}
We proceed to show that the Wilson loop average in Yang-Mills theory is rewritten into a bosonic string theory.  
We see that
\begin{equation}
\int d^4x (\epsilon_{\mu\nu\rho\sigma}\partial^\nu B^{\rho\sigma})^2
= \int d^4x 2(-B_{\mu\nu} \partial^2 B_{\mu\nu} - B_{\mu\nu} \partial_\nu \partial_\lambda B_{\lambda\mu} + B_{\nu\mu} \partial_\mu \partial_\lambda B_{\lambda\nu})  
\end{equation}
using the integration by parts. 
There are two ways to perform the integration over $B_{\mu\nu}$ field to obtain the Wilson loop average.
One way is to identify the theory $S_d[B]$ written in terms of $B$ with a gauge-fixed version of a gauge-invariant master theory with an action $S_M[B,\Lambda]$, see Appendix B.  
This way is applicable to the case up to $\ell=0$.
In this case, the gauge fixing condition for $B$ can be chosen such that
$\partial_\rho B_{\rho\sigma}=0$ which greatly simplifies the calculation, see Appendix \ref{sec:KRfield} (or Appendix D of \cite{Kondo00}).
Another way is to deal directly with the action $S_d[B]$ without imposing any condition on $B$.  
This way is applicable to any $\ell$.  In this case, the Wilson loop average has extra contributions corresponding to the boundary terms in the coordinate representation.  However, it turns out that the boundary term is not responsible for the area decay, see Appendix \ref{sec:Binteg}.  Therefore, we can neglect the boundary term to obtain the area decay of the Wilson loop average.  
\par  
Thus,  the relevant part of the action responsible for the area decay of the Wilson loop average reads
\begin{align}
  S_d'[B] = \int d^4x  {1 \over 4\tilde{\kappa} M^2} 
  B_{\mu\nu}  I_{\mu\nu\rho\sigma}
   \left[   M^2 - \partial^2  
+  {1 \over \tilde{M}^2} (\partial^2)^2 
  \right] B_{\rho\sigma}  + \cdots ,  
  \label{Bth2}
\end{align}
where
\begin{align} 
 M^2 :=  {4 \kappa_0/\tilde{\kappa} \over K^{(0)} \sigma_{vc}+{1 \over 4}\omega^{(1)}} ,
\quad 
\tilde{M}^2 := 2 \tilde{\kappa} {K^{(0)} \sigma_{vc}+{1 \over 4}\omega^{(1)} \over K^{(1)}} ,
\quad
\tilde{\kappa} :=  (1/\kappa_0 + 1/\kappa_0^2)^{-1} .
\label{M}
\end{align}

\par
The general form of the full gluon propagator in momentum space is given by
\begin{align}
  \tilde{\mathscr{D}}_{\mu\nu}^{AB}(p) 
=&  \delta^{AB}[(\delta_{\mu\nu}p^2-p_\mu p_\nu)A(p)+\lambda^{-1} p_\mu p_\nu B(p) + \delta_{\mu\nu} C(p)]^{-1} 
\\
=&  \delta^{AB} \left[ {1 \over p^2 A(p)+C(p)} P_{\mu\nu}^T + {\lambda \over p^2B(p)+\lambda C(p)} P_{\mu\nu}^L \right] ,
\end{align}
where 
$
 P_{\mu\nu}^T := \delta_{\mu\nu} - p_\mu p_\nu/p^2 
$
and
$
 P_{\mu\nu}^L := p_\mu p_\nu/p^2.
$
The asymptotic form of the full propagator in the large $p$ region reads in the Landau gauge
\begin{align}
  \tilde{\mathscr{D}}_{\mu\nu}^{AB}(p) 
=  \delta^{AB} \left[ {1 \over p^2 A(p)} - {C(p) \over p^4 A^2(p)} +O(1/p^6) \right] P_{\mu\nu}^T   ,
\label{largep}
\end{align}
while in the small $p$ region 
\begin{align}
\tilde{\mathscr{D}}_{\mu\nu}^{AB}(p) 
=  \delta^{AB} \left[ {1 \over C(p)} - {p^2 A(p) \over C^2(p)} + O(p^4) \right] P_{\mu\nu}^T   .
\label{smallp}
\end{align}
\par
The OPE result is inconsistent with the absence of both current and dynamical gluon mass, i.e., $C(p) \equiv 0$.
In fact, the OPE calculation with the renormalization group improvement (\ref{gluonOPE}) in the Landau gauge $\lambda=0$ yields
\begin{align}
  \tilde{\mathscr{D}}_{\mu\nu}^{AB}(p) 
=&   \tilde{D}_{\mu\nu}^{AB}(p)  + \tilde{W}^{[{1 \over 2}\mathscr{A}\mathscr{A}]}{}_{\mu\nu}^{AB}(p) \langle {1 \over 2}\mathscr{A}_\rho \cdot \mathscr{A}_\rho\left(X \right)  \rangle   
\\
=&  \delta^{AB} \left[ {-1 \over p^2}
 \left({\ln p/\Lambda_0 \over \ln \mu/\Lambda_0} \right)^{-{13 \over 6}{N_c \over \beta_0}} 
+ {N_c g^2 \over 2(N_c^2-1)} {\langle {1 \over 2}\mathscr{A}^2 \rangle \over p^4} 
 \left({\ln p/\Lambda_0 \over \ln \mu/\Lambda_0} \right)^{{3\over 4}{N_c \over \beta_0}} + O(1/p^6)
\right] P_{\mu\nu}^T   .
\end{align}
From the consistency of the OPE result with the general asymptotic behavior (\ref{largep}), we are lead to the existence of non-vanishing gluon mass function $C$ with the logarithmic behavior:
\begin{equation}
  A(p) = \left({\ln p/\Lambda_0 \over \ln \mu/\Lambda_0} \right)^{{13 \over 6}{N_c \over \beta_0}} ,
\quad
C(p) =   {N_c g^2 \over 2(N_c^2-1)} \langle {1 \over 2}\mathscr{A}^2 \rangle  
 \left({\ln p/\Lambda_0 \over \ln \mu/\Lambda_0} \right)^{{61 \over 12}{N_c \over \beta_0}}  .
\label{massf}
\end{equation}

On the other hand, we find
\begin{align}
 K_{\mu\nu}^{(0)} \sigma_{vc} =& a_0 \int {d^4p \over (2\pi)^4} \  
  \tilde{G}^{AB}(-p)  \tilde{W}^{[{1 \over 2}\mathscr{A}\mathscr{A}]}{}_{\mu\nu}^{AB}(p) 
\langle {1 \over 2}\mathscr{A}_\rho \cdot \mathscr{A}_\rho\left(X \right)  \rangle 
\nonumber\\
=& a_0 \int {d^4p \over (2\pi)^4} \    \tilde{G}^{AB}(-p) 
 [\tilde{\mathscr{D}}_{\mu\nu}^{AB}(p) -  \tilde{D}_{\mu\nu}^{AB}(p) ] ,
\label{defK2}
\end{align}
By substituting the asymptotic form (\ref{smallp}) into (\ref{defK2}), we obtain in the Landau gauge $\lambda=0$
\begin{align}
 K_{\mu\nu}^{(0)} \sigma_{vc} =& a_0 \int {d^4p \over (2\pi)^4} \    \tilde{G}^{AB}(-p) \delta^{AB} \left\{
  \left[ {1 \over C(p)} - {p^2 A(p) \over C^2(p)} + O(p^4) \right] - p^2 \right\} P_{\mu\nu}^T 
\nonumber\\
=&  \left[ {N_c g^2 \over 2(N_c^2-1)} \langle {1 \over 2}\mathscr{A}^2 \rangle \right]^{-1} a_0 \int {d^4p \over (2\pi)^4} \
\tilde{G}^{AB}(-p) \delta^{AB} \
\left\{
 \left({\ln p/\Lambda_0 \over \ln \mu/\Lambda_0} \right)^{\#} 
+ O(p^2)  
 \right\} P_{\mu\nu}^T  ,
\label{defK3}
\end{align}
where $\tilde{G}^{AB}(p)$ is the Fourier transform of $G^{AB}(x-y)$ defined by (\ref{G}) and $\#$ is a numerical number to be calculated. The integration over four momenta in (\ref{defK3}) gives a dimensionless number. 
Therefore, $K^{(0)} \sigma_{vc}$ is proportional to the inverse of the vacuum condensate,%
\footnote{
Of course, it is possible to perform the loop calculation to obtain the coefficient.  
The RG improved OPE (\ref{massf}) leads to  
$\#=-{61 \over 12}{N_c \over \beta_0}$, as shown above.  
However, in order to obtain reliable results over the whole momentum region, it is desirable to solve the flow equation of the non-perturbative renormalization group.  The relevant results from this viewpoint will be reported in forthcoming papers \cite{IKKMS02}.
} 
\begin{equation}
 K^{(0)} \sigma_{vc} \cong 
  \left[ {N_c g^2 \over 2(N_c^2-1)} \langle {1 \over 2}\mathscr{A}^2 \rangle \right]^{-1}   .
\end{equation}
In the absence of gluon pair condensation
$\langle \mathscr{A}^2 \rangle=0$, $M$ defined by (\ref{M}) vanishes, $M=0$.  Roughly speaking, $M^2$ is proportional to the gluon condensation of mass dimension 2,
i.e., apart from a numerical coefficient
\begin{equation}
 M^2 \cong g^2 \langle {1 \over 2}\mathscr{A}_\mu \mathscr{A}_\mu  \rangle = g^2 \sigma_{vc} .
\end{equation}

\par
In this paper we have taken into account only the first two terms in (\ref{expan}) corresponding to $\ell=0$ and $\ell=1$.
An advantage of including the $\ell=1$ term in addition to $\ell=0$ one is that the ultraviolet (UV) cutoff $\Lambda$ can be removed in the final stage of calculating the string tension to obtain a finite value for the string tension (See Appendix \ref{sec:Wilsonloop}).   
\par
The integration over the $B_{\mu\nu}$ field is achieved by Gaussian integration to yield 
\begin{align}
   W(C) 
  \cong& \exp \left[ -g^2 {J^2 \over 4} \tilde{\kappa} \chi 
 \sum_{i=1,2} (-1)^{i+1} (\Theta, I(-\partial^2 +M_i^2)^{-1} \Theta) 
+ {\kappa_0 \over 24} (gJ)^2  (\Theta, I \Theta)
\right] ,
\label{W}
\end{align}
where we have defined
\footnote{
We have defined the $L^2$ inner product:
\begin{equation}
  (\Theta, I K \Theta)
:= \int d^4x \int d^4y \Theta_{\mu\nu}(x) I_{\mu\nu\rho\sigma} K(x,y) \Theta_{\rho\sigma}(y) .
\end{equation}
In particular,
\begin{equation}
  (\Theta, I \Theta)
:= \int d^4x \Theta_{\mu\nu}(x) I_{\mu\nu\rho\sigma} \Theta_{\rho\sigma}(x) .
\end{equation}
}
\begin{align}
 M_{1,2}^2 :=& {1 \over 2} \tilde{M}^2 \left(1 \mp 
\sqrt{1-4{M^2 \over \tilde{M}^2}} \right)
\quad (M_1<M_2),
\\
 \chi :=& M_1^2 M_2^2/(M_2^2-M_1^2) 
= M^2/\sqrt{1-4{M^2 \over \tilde{M}^2}}.
\end{align}
We mention two special cases.
We consider the first limit (apparent London limit): $M \ll \tilde{M} \rightarrow \infty$, namely, 
$M_2 \cong \tilde{M} \rightarrow \infty$ and $M_1 \cong M < \infty$.
In this case, the mode with a huge mass $M_2$ decouples from the theory and we have only to include the mode with a mass $M_1$ alone corresponding to $\ell=0$.
On the other hand, in the second limit (apparent Bogomol'nyi limit) defined by $M_1 \cong M_2$, namely, $M_1 \cong M_2 \cong \tilde{M}/\sqrt{2} \cong \sqrt{2}M$, 
we must include the second term too, i.e., $\ell=0,1$.
\par 
\begin{figure}[htbp]
\begin{center}
\[
\begin{array}{cl}
\begin{array}{c}
\includegraphics{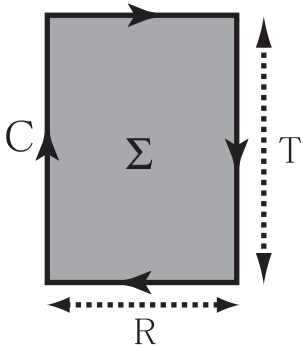}
\end{array}
\end{array}
\]
\caption{A planar Wilson loop with side lengths $R$ and $T$.}
\label{fig:Wilsonloop2}
\end{center}
\end{figure}

\par
Now we proceed to calculate the Wilson loop average.  We choose a rectangular loop $C$ with side lengths $R$ and $T$ (Fig.~\ref{fig:Wilsonloop2})  for simplifying the calculation.  
It is shown by explicitly calculating the integral in the argument of the exponential (\ref{W})  (see Appendix \ref{sec:Wilsonloop} or Appendix F of \cite{Kondo00}) that the Wilson loop average exhibits the area law decay :
\begin{equation}
  W(C) \cong \exp [ - \sigma_{st} \text{Area}(C) ] ,
  \quad 
  \sigma_{st} = J^2 g^2 \tilde{\kappa} {\chi \over 8\pi} \ln {M_2 \over M_1} ,
\end{equation}
where $\text{Area}(C)=RT$ is the minimal area of the surface spanned by the rectangular loop $C$ and $\sigma_{st}$ is the string tension.
The string tension in the first limit reads
$
  \sigma_{st} = J^2 g^2 \tilde{\kappa} {M^2 \over 4\pi} \ln {\tilde M \over M} ,
$
while in the second limit 
$
  \sigma_{st} = J^2 g^2 M^2 .
$
What the area law holds irrespective of the shape of the loop $C$ can be understood through the string representation given below.

\subsection{Step 8: String representation}

\begin{figure}[htbp]
\begin{center}
\[
\begin{array}{cl}
\begin{array}{c}
\includegraphics{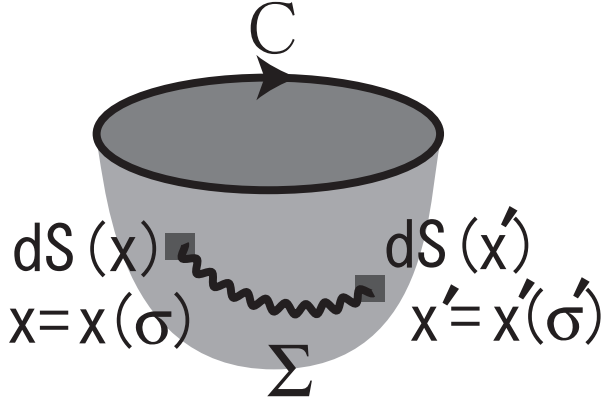}
\end{array}
\end{array}
\]
\caption{Interacting two surface elements $dS^{\mu\nu}(x)$ and $dS^{\rho\sigma}(x')$ in the surface $\Sigma$ spanned by the Wilson loop $C$.}
\label{fig:stringrep}
\end{center}
\end{figure}

\par
Now we obtain the string representation of the Wilson loop average in Yang-Mills theory.
Let $\sigma=(\sigma^1,\sigma^2)$ be a two-dimensional coordinate on the world sheet, while the target space coordinates is expressed by $x_\mu=x_\mu(\sigma)$.  Then the infinitesimal surface element,
\begin{equation}
  dS_{\mu\nu}(x(\sigma))=\sqrt{g(\sigma)}t_{\mu\nu}(\sigma) d^2\sigma ,
\end{equation}
is expressed by 
the so-called {\it extrinsic curvature} tensor of the surface,
\begin{equation}
  t_{\mu\nu}(\sigma)={\epsilon^{ab} \over \sqrt{g(\sigma)}} \partial_a x_\mu(\sigma) \partial_b x_\nu(\sigma) ,
\end{equation}
where the determinant $g(\sigma)=\det||g_{ab}(\sigma)||$ is calculated from the induced metric tensor of the surface defined by
$
 g_{ab}(\sigma)=\partial_a x_\mu(\sigma) \partial_b x_\mu(\sigma)
$
with the derivative,
$
 \partial_a = {\partial \over \partial \sigma^a} \quad (a=1,2) .
$ 
We use the conformal gauge for the induced metric 
$g_{ab}(\sigma)=\sqrt{g(\sigma)}\delta^{ab}$.
\par
We can observe in the expression (\ref{W}) that there are massive propagating modes mediating the interaction between two surface elements $dS^{\mu\nu}(x)$ and $dS^{\rho\sigma}(x')$ in the surface $\Sigma$ spanned by the Wilson loop $C$, see Fig.~\ref{fig:stringrep}.
It is shown that the {\it derivative expansion} in powers of 
$\zeta_i^a := (\sigma'-\sigma)^aM_i$ leads to
\begin{align}
 & (\Theta, I(-\partial^2 +M_i^2)^{-1} \Theta) 
  \nonumber\\
  =&  \int_{\Sigma} d^2S^{\mu\nu}(x(\sigma)) \int_{S} d^2S^{\rho\sigma}(x'(\sigma')) I_{\mu\nu\rho\sigma}  (-\partial^2 +M_i^2)^{-1}(x(\sigma),x'(\sigma')) 
  \nonumber\\
  =& \int_{\Sigma} d^2\sigma \sqrt{g} \left[ 4 M_i^{-2} \mathscr{M}_i^{(0)} - {1 \over 4} M_i^{-4} \mathscr{M}_i^{(2)} g^{ab}(\partial_a t_{\mu\nu})(\partial_b t_{\mu\nu}) \right] + O( \zeta^{4}) ,
\end{align}
where $\mathscr{M}_i^{(n)}$ is the $n$-th moment defined by
$
 \mathscr{M}_i^{(n)} := \int d^2z (z_i^2)^{n} G_{M_i}(z^2)$ for $z_i^a := g^{1/4} \zeta_i^a ,
$
from the Green function
$
 G_M(z^2) =(-\partial^2+M^2)^{-1}(x,x') = {1 \over 4\pi^2}{M \over |x-x'|}K_1(M|x-x'|) ,
$
with $K_1$ being the modified Bessel function, see Appendix \ref{sec:dexp}.  

Thus we obtain the confining string theory for the Wilson loop average:
\begin{align}
  W(C) \cong&  \exp (-S_{cs}[x]) ,
  \nonumber\\
    S_{cs} =& \sigma_{st} \int_{\Sigma} d^2\sigma \sqrt{g} 
  + \alpha_0^{-1} \int_{\Sigma} d^2\sigma \sqrt{g}  g^{ab} \partial_a t_{\mu\nu} \partial_b t_{\mu\nu} + \kappa_t \int_{\Sigma} d^2\sigma \sqrt{g}R + \cdots ,
\end{align}
where%
\footnote{
Here we have fixed a surface spanned by the Wilson loop.
Taking into account the summation over the possible surfaces, we obtain
$
 W(C) \cong \int \mathcal{D}x_\mu(\sigma) J[x] \exp (-S_{cs}[x])
$ 
with $J[x]$ being the Jacobian associated with the change of variables from the vorticity tensor to the target space coordinates 
(see e.g., \cite{ACPZ96,PS91}).
}
the first term is the Nambu-Goto action, the second term is the rigidity term, and 
the third term is the intrinsic curvature term and 
the parameters are determined as
\begin{subequations}
\begin{align}
 \sigma_{st}  =& {J^2g^2  \over 4\pi} \tilde{\kappa} \chi \ln \left( {M_2 \over M_1} \right) ,
 \\
 \alpha_0^{-1} =& - J^2g^2 \tilde{\kappa} {1 \over 4\pi}  <0  ,
 \\
 \kappa_t =& {2 \over 3} J^2g^2 \tilde{\kappa} {1 \over 4\pi}  > 0  .
\end{align}
\end{subequations}
The details of the calculation is given in Appendix \ref{sec:dexp} (see also Appendix G of \cite{Kondo00}) which could be compared with a review article \cite{AES96,Antonov99} where a similar calculations were performed. 
The string theory just derived is nothing but the rigid string with a negative rigidity term proposed by Polyakov \cite{Polyakov96}.
(See \cite{Polyakov86,Kleinert86} and \cite{BPT87,OY87,BZ87,DG88,PY92,KC96,DQT96,DT97} for the detailed studies of the rigid string.) 
In the absence of the vacuum condensate of mass dimension 2, i.e., $\sigma_{vc} \rightarrow 0$, 
the string tension vanishes $\sigma_{st} \rightarrow 0$ and the area law decay is lost. 
In other words, non-zero vacuum condensation $\sigma_{vc}\not=0$ is suggested from the non-zero string tension $\sigma_{st}\not=0$. 
\par

\subsection{Step 9: Dual Ginzburg-Landau theory}
We proceed to show that the gluon condensate of mass dimension 2 is equivalent to the monopole condensation which is believed to be a mechanism for the dual superconductivity of gluodynamics.%
\footnote{
In the previous paper \cite{KondoI}, we stressed the importance of the term: 
\begin{equation}
  \int d^4x h_{\mu\nu} f_{\mu\nu}^V 
  = \int d^4x B_{\mu\nu} {}^*f_{\mu\nu}^V 
  = \int d^4x \partial_\mu b_\nu  {}^*f_{\mu\nu}^V 
  = \int d^4x b_\mu k_\mu ,
\end{equation}
where we have defined the monopole current:
$
 k_\mu = \partial_\nu  {}^*f_{\mu\nu}^V  
$
after the Hodge decomposition of $B_{\mu\nu}=\partial_\mu b_\nu - \partial_\nu b_\mu + \cdots$. 
The VEV of the quadratic term yields the mass term of the dual (Abelian) gauge field $b_\mu$, if
$\langle \langle k_\mu(x) k_\nu(y) \rangle_{V} \rangle_{YM}  
=g^2 \delta_{\mu\nu}\delta^4(x-y) K+ \cdots$ as
\begin{equation}
 \Big\langle \Big\langle \left( \int d^4x h_{\mu\nu} f_{\mu\nu}^V \right)^2 \Big\rangle_{V} \Big\rangle_{YM}  
   = \int d^4x \int d^4y b_\mu(x) b_\nu(y) \langle \langle k_\mu(x) k_\nu(y) \rangle_{V} \rangle_{YM}  
   \rightarrow \int d^4x {1 \over 2}m_b^2 U_\mu^2(x) ,
\end{equation}
where $m_b^2$ is proportional to $K$. 
This fact leads to an interpretation that the dual gauge field becomes massive due to the dual Meissner effect caused by monopole condensation 
$\langle \langle k_\mu(x) k_\mu(x) \rangle_{V} \rangle_{YM}  \not=0$.
}
The dual transformation (or Fourier transformation) can be used to transform the massive antisymmetric tensor field theory into another equivalent theory.  
It is well known \cite{Kawai81,Orland82} that in $D$-dimensional space-time, a massless antisymmetric tensor field of rank $p$ ($p$-form) is dual to the $(D-p-2)$-form, while a massive antisymmetric field of rank $p$ is dual to the $(D-p-1)$-form.  
In fact, it is shown in Appendix \ref{sec:dual} that the effective theory (\ref{Bth}) retained up to the quadratic terms in the massive Kalb-Ramond \cite{KR74} field $B_{\mu\nu}$ is dual to the massive vector field $U_\mu$ theory described by the dual Ginzburg-Landau (DGL) theory in the London limit:
\begin{align}
  S_{DGL} = \int d^4x \left[ {\tilde{\kappa} \over 4} (U_{\mu\nu}+U_{\mu\nu}^S)^2 + {1 \over 2}m_U^2 U_\mu U_\mu \right] ,
\end{align}
where $U_{\mu\nu}:=\partial_\mu U_\nu - \partial_\nu U_\mu$,
$U_{\mu\nu}^S$ is the Dirac string tensor defined by
$U_{\mu\nu}^S(x) := Jg  \Theta_{\mu\nu}(x)$
and
the radial mode of the scalar field is frozen and the phase is absorbed into the vector field $U_\mu$ to make the vector field massive with the mass
\begin{equation}
  m_U  = \tilde{\kappa}^{1/2} M \cong 
\sqrt{  g^2 \langle \mathscr{A}^2 \rangle } .
\end{equation}
Here the kinetic term of $B_{\mu\nu}$ corresponds to the mass term of $U_\mu$ and vice versa.
\par
In the limit $M \rightarrow 0$, the dual gauge mass vanishes $m_U \rightarrow 0$.  In this limit, if the field $U_\mu$ is rescaled as $U_\mu \rightarrow U_\mu/m_U$, then the configuration $U_{\mu\nu}=0$ becomes dominant, so that the vector theory written in terms of $U_\mu=\partial_\mu \phi$  reduces to a scalar field theory written in terms of $\phi$.
Therefore, for the dual Ginzburg-Landau theory to describe the massive dual vector as a consequence of the dual Meissner effect, the dynamical generation of the kinetic term of $B_{\mu\nu}$ is indispensable through non-zero value of $\sigma_{vc}$.  
In other words,  the mass gap caused by the gluon (and ghost) pair condensation in the original Yang-Mills theory is equivalent to the dual Meissner effect caused by monopole condensation in the dual theory.

\par
As has been shown, the London limit (an extreme limit of the type II dual superconductor) of the DGL theory is obtained by including the quadratic term in the effective theory.  
In order to reproduce the DGL theory beyond the London limit, it is necessary to include the quartic term in $h$, see Appendix \ref{sec:dual}.

The final result  still depends on $\kappa$.  
It should be remarked that the original theory is independent of $\kappa$.  Therefore, if we can take into account all the terms of the cumulant expansion, the result should be $\kappa$-independent.  However, the dual theory obtained from the truncated effective action may depend on $\kappa$.  We hope that the dependence will be substantially small.  If so, the truncation of the infinite series turns out to be a good approximation.  
The precise estimation of this dependence will be given in a forthcoming paper based on the exact renormalization group method \cite{IKKMS02}.  

\section{Remarks}

We can consider an alternative step for Step 5.
This enables us to discuss the relationship between the gauge potential correlator $\mathscr{A}-\mathscr{A}$  and the  field strength correlator $\mathcal{F}-\mathcal{F}$ briefly.
The effective theory $S_d[h]$ is also obtained in connection with $\mathcal{F}-\mathcal{F}$ correlation function.
In fact,  $f_{\mu\nu}^{V}$ has also the manifestly gauge- and BRST-invariant form: 
\begin{align}
  f_{\mu\nu}^{V}(x) :=  {\bf V}(x) \cdot \mathcal{F}_{\mu\nu}(x) 
- g^{-1} {\bf V}(x) \cdot (D_\mu{\bf V}(x) \times D_\nu{\bf V}(x))   ,
\end{align}
where $D_\mu  := \partial_\mu  -ig [\mathscr{A}_\mu, \cdot ]$.
It is easy to see that the first term 
${\bf V}(x) \cdot \mathcal{F}_{\mu\nu}(x)$
and the second term 
$
\varOmega_{\mu\nu}(x) := {\bf V}(x) \cdot (D_\mu{\bf V}(x) \times D_\nu{\bf V}(x))
 = - \varOmega_{\nu\mu}(x)  
$
of RHS of this equation is separately gauge invariant.
Here note that 
$
 \langle \mathcal{F}_{\mu\nu}^A(x) \mathcal{F}_{\rho\sigma}^B(y) \rangle_{YM}
$ 
alone is not gauge invariant.
Therefore, using this expression in (\ref{dcex}), 
we obtain an alternative BRST-invariant action:
\begin{align}
  S_h =& \int d^4x  {1 \over 4\kappa} h_{\mu\nu}(x)h_{\mu\nu}(x) 
-  {1 \over 8\kappa^2}  \int d^4x \int d^4y 
h_{\mu\nu}(x) h_{\rho\sigma}(y)
\langle \langle f^{V}_{\mu\nu}(x)  f^{V}_{\rho\sigma}(y) \rangle_{V}^{\text{con}} \rangle_{YM}^{\text{con}} 
+ \cdots 
\nonumber\\
=& \int d^4x  {1 \over 4\kappa} h_{\mu\nu}(x)h_{\mu\nu}(x) 
-  {1 \over 8\kappa^2} g^{-2} \int d^4x \int d^4y 
h_{\mu\nu}(x) h_{\rho\sigma}(y)
\langle \langle \varOmega_{\mu\nu}(x)  \varOmega_{\rho\sigma}(y) \rangle_{V}^{\text{con}} \rangle_{YM}^{\text{con}} 
  \nonumber\\&
  -  {1 \over 8\kappa^2}  \int d^4x \int d^4 y
 \langle V^A(x) V^B(y) \rangle_{V}^{\text{con}} \langle \mathcal{F}_{\mu\nu}^A(x) \mathcal{F}_{\rho\sigma}^B(y) \rangle_{YM}^{\text{con}} 
   h_{\mu\nu}(x)  h_{\rho\sigma}(y) 
+ \cdots .
 \label{nlaction2}
\end{align}
By expanding $h(x)$ and $h(y)$ in the formal power series in $r:=x-y$ around the coordinate $X:=(x+y)/2$, 
the above action is cast into
\begin{align}
  S_h =  \int d^4X \left[ {1 \over 4\kappa} \left( I_{\mu\nu\rho\sigma}-{1 \over 2\kappa} K^{(0)}_{\mu\nu\rho\sigma}\right) h_{\mu\nu}(X)h_{\rho\sigma}(X) 
-  {K^{(1)}_{\mu\nu\rho\sigma} \over 64\kappa^2}  
  h_{\mu\nu}(X) \partial^2 h_{\rho\sigma}(X) 
 \right] + \cdots   ,
 \label{nlaction3}
\end{align}
where we have assumed the translational invariance for 
the correlation function as 
 functions  of $r^2=(x-y)^2$
and we have defined
\begin{align}
 K^{(\ell)}_{\mu\nu\rho\sigma} := \int d^4(x-y) \ \{(x-y)^2\}^\ell  
\langle \langle f^{V}_{\mu\nu}(x)  f^{V}_{\rho\sigma}(y) \rangle_{V}^{\text{con}} \rangle_{YM}^{\text{con}}   .
\label{cumu2}
\end{align}
\par
This effective action (\ref{nlaction2}) just derived should be compared with the previous one (\ref{nlaction}).
An advantage of this approach is that the equation preserves the manifest gauge and BRST invariance.
The disadvantages are as follows.   
The evaluation of the field-strength correlator (e.g., 
$\langle \mathcal{F}_{\mu\nu}^A(x) \mathcal{F}_{\rho\sigma}^B(y) \rangle_{YM}^{\text{con}}$)
in the effective action (\ref{nlaction2})
is much harder than the gluon field correlator 
$\langle \mathscr{A}_\mu^A(x) \mathscr{A}_\nu^B(y) \rangle_{YM}^{\text{con}}$
in (\ref{nlaction}).
Furthermore, it is difficult to relate this version of the effective theory with the gluon condensation of dimension 4, i.e., 
$\langle \mathcal{F}_{\mu\nu}^A(x) \mathcal{F}_{\mu\nu}^A(x) \rangle_{YM} $, since the OPE of the composite operator $\mathcal{F}_{\mu\nu}^A(x) \mathcal{F}_{\rho\sigma}^B(y)$ does not include the gluon condensation of mass dimension 4 and hence cannot give the manifestly gauge and BRST invariant result, see \cite{LO92}.
However, the following identification of the gauge-invariant correlation function might be related to the approach of the stochastic vacuum model (SVM).%
\footnote{In the non-perturbative study of QCD, the cumulant expansion is extensively utilized by the SVM \cite{Dosch87} where the different version of the non-Abelian Stokes theorem is adopted.
In the SVM, the approximation of neglecting higher order cumulants is called the bilocal approximation. 
The Gaussian correlator has been calculated on a lattice \cite{GP92,BBDV98,BBV98} and 
the validity of bilocal approximation in SVM was confirmed by Monte Carlo simulation on a lattice \cite{BBV98}.
The relationship between two gluon correlators and the dual field correlator in the supposed DAH model has been investigated in \cite{BBDV98}.
The authors would like to thank Dmitri Antonov and Nora Brambilla for this information.
} 
\begin{align}
   \langle \langle f^{V}_{\mu\nu}(x)  f^{V}_{\rho\sigma}(y) \rangle_{V}^{\text{con}} \rangle_{YM}^{\text{con}}
   =&   I_{\mu\nu\rho\sigma}D_1((x-y)^2) 
 +   [ \partial_\mu ((x-y)_\rho \delta_{\nu\sigma}-(x-y)_\sigma \delta_{\nu\rho}) 
  \nonumber\\
  &+ \partial_\nu ((x-y)_\sigma \delta_{\mu\rho}-(x-y)_\rho \delta_{\mu\sigma})  ] D_2((x-y)^2) .
\end{align}
Once the two functions $D_1$ and $D_2$ are determined, it is possible to calculate $K^{(\ell)}_{\mu\nu\rho\sigma}$ of (\ref{cumu2}) to fix the action of the effective theory (\ref{nlaction3}).
\par
In the Fock-Schwinger (FS) gauge $x^\mu \mathscr{A}_\mu(x)=0$ (or coordinate gauge) 
\cite{PT84,Shifman80,NSVZ85,ESS88}, it is possible to compare both approaches, since the gauge field $\mathscr{A}_\mu$ is written in terms of the field strength $\mathcal{F}_{\mu\nu}$ in the FS gauge:
\begin{align}
  \mathscr{A}_\mu(x) =& - \int_0^1 da \ ax^\nu \mathcal{F}_{\mu\nu}(ax) 
\nonumber\\
  =& - \sum_{n=0}^{\infty} {1 \over (n+2)n!} x^\nu x^{\omega_1} x^{\omega_2} \cdots x^{\omega_n} [D_{\omega_1}(0), [D_{\omega_2}(0), [ \cdots [D_{\omega_n}(0), \mathcal{F}_{\mu\nu}  ] \cdots  ]]]
\nonumber\\
  =& - {1 \over 2} \mathcal{F}_{\mu\nu}(0) x^\nu 
  - {1 \over 3} x^\nu x^\lambda [D_\lambda(0), \mathcal{F}_{\mu\nu} (0)]
  - {1 \over 8} x^\nu x^\lambda x^\tau [D_\lambda(0), [D_\tau(0), \mathcal{F}_{\mu\nu}(0)] + \cdots ,
\end{align}
where 
$D_\mu(0) := \partial_\mu - ig [\mathscr{A}_\mu(0), \cdot]$. 
This implies the gluon propagator 
\begin{align}
  \langle \mathscr{A}_\mu^A(x) \mathscr{A}_\nu^B(y) \rangle
  =& {1 \over 4}x^\rho y^\sigma \langle \mathcal{F}_{\mu\rho}^A(0) \mathcal{F}_{\nu\sigma}^B(0) \rangle + \cdots 
  \nonumber\\
  =& {1 \over 4}x^\rho y^\sigma  {\delta^{AB}\langle \mathcal{F}_{\alpha\beta}(0) \cdot \mathcal{F}_{\alpha\beta}(0) \rangle \over (N^2-1)D(D-1)} (\delta_{\mu\nu}\delta_{\rho\sigma}-\delta_{\mu\sigma}\delta_{\rho\nu}) + \cdots .
\end{align}
Hence, two types of gluon condensations are related as
\begin{align}
  \langle \mathscr{A}_\mu(x) \cdot \mathscr{A}_\mu(x) \rangle
  =  {1 \over 4D}  \langle \mathcal{F}_{\alpha\beta}(0) \cdot \mathcal{F}_{\alpha\beta}(0) \rangle x^2  + \cdots .
  \label{conden}
\end{align}
It should be remarked that the FS gauge does not retain the translational invariance.
In fact, we do not assume translational invariance from the beginning in this paper.\footnote{
The authors in \cite{SY99} have concluded that the condensate
$\langle \mathscr{A}_\mu^2 \rangle$ is zero, provided that the translational invariance holds and the approximation of the vacuum dominance in intermediate state is accepted.  
Therefore, our results suggest that this approximation is not so good.  
Contrary to the claim in \cite{SY99}, the vanishing of the condensate is not essential to retain the gauge invariance of the vacuum energy.
}
The translational invariance implies that the VEV of $\mathcal{O}$ is equal to the VEV of the integrand $Q(x)$:
\begin{align}
  \langle \mathcal{O} \rangle 
= \Omega^{-1} \int d^4x \langle Q(x) \rangle 
= \Omega^{-1} \int d^4x \langle Q(0) \rangle 
= \langle Q(0) \rangle .
\end{align}
In the translationally non-invariant gauge, the situation is quite different.  
In the FS gauge, the vacuum condensate of mass dimension 2 is related to that of mass dimension 4 as
\begin{align}
  \langle \mathcal{O} \rangle 
= \Omega^{-1} \int d^4x \left\langle {1 \over 2}\mathscr{A}_\mu(x) \cdot \mathscr{A}_\mu(x) \right\rangle 
= {1 \over 8D}   \langle \mathcal{F}_{\alpha\beta}(0) \cdot \mathcal{F}_{\alpha\beta}(0)   \rangle 
\Omega^{-1} \int d^4x \ x^2  + \cdots  .
\end{align}
In order to make this quantity well-defined, it is necessary to introduce the long-distance or IR cutoff.%
\footnote{
If the theory is defined on a finite space-time volume, then $\langle \mathcal{O} \rangle$ is automatically finite.  However, it does not guarantee the existence of the infinite volume limit.    
}
\par
In the Lorentz gauge, defining 
\begin{equation}
   \tilde{\mathcal{O}}(q)  
  = \Omega^{-1} \int d^4x \ e^{iqx}
  \left[ {1 \over 2} \mathscr{A}_\mu(x) \mathscr{A}_\mu(x) + \lambda i \bar{\mathscr{C}}(x) \mathscr{C}(x) \right] ,
\end{equation}
we find that the vacuum condensate of mass dimension 2 proposed in \cite{Kondo01} is equivalent to the statement for condensation of the zero-momentum mode
$
 \langle \mathcal{O} \rangle
 = \langle \tilde{\mathcal{O}}(q=0) \rangle
 \not= 0
 $.
Therefore, this can be interpreted as the simultaneous Bose-Einstein condensation of gluon pair and ghost-antighost pair.
The phenomenological description of gluon pair condensation has been discussed in \cite{CS86,CS84,GJJS86}.
  This issue is to be investigated subsequently in detail.

\section{Conclusion and discussion}

In this paper we have proposed a strategy to obtain the effective field theory or the effective string theory of the Yang-Mills theory with an insertion of the Wilson loop operator.
In fact, we have obtained a number of effective theories which are equivalent to the original Yang-Mills theory within the validity of the approximations adopted here (weak-field approximation and lower derivative expansion).  
\par
The first effective theory is written in terms of an antisymmetric tensor field $h$ which is coupled with the surface (vorticity tensor) spanned by the Wilson loop $C$.  At first, the antisymmetric tensor $h$ was introduced as an auxiliary field without its kinetic term.  After integrating out all the other fundamental fields which are expected to become massive,  $h$ has acquired the kinetic term  
 due to the gluon pair condensation in the Landau gauge 
(a special case of the simultaneous Bose-Einstein condensation of gluon pairs and ghost-antighost pairs in the general Lorentz gauge).  
 This effect has been estimated as a non-vanishing vacuum condensate of a composite operator with mass dimension 2 which is a color-singlet quantity and a BRST invariant combination of the gluon field and the ghost (antighost) field \cite{Kondo01}. 
\par
The second effective theory is a confining string theory with a rigidity term, which was derived through the non-local action for the vorticity current tensor.  The bosonic string theory known as a rigid string was conjectured to be an effective theory of gluodynamics  a long time ago.  
At least, in the weak-field approximation and the derivative expansion, our result confirms this conjecture.%
\footnote{
In the modified MA gauge, all order contributions from $\Omega_{\mu\nu}$ were incorporated by making use of the dimensional reduction to the two-dimensional nonlinear sigma model and instanton calculus in the dilute gas approximation within the framework of a deformation of a topological field theory  
as a reformulation of Yang-Mills theory \cite{KondoII}.
}  
\par
The third effective theory is the dual Ginzburg-Landau theory obtained through the duality transformation in the sense of the electric-magnetic duality.   
The mass gap in the dual theory is understood as a manifestation of the dual Meissner effect caused by the monopole condensation.  
\par
An advantage of our approach is that we must assume from the beginning neither the existence of mass gap nor a bare mass term in Yang-Mills theory.  Such assumptions were often adopted to study the non-perturbative feature of Yang-Mills theory.  Just as the dynamical generation of quark mass associated with the spontaneous breaking of chiral symmetry, the gluon and ghost masses are of dynamical origin (namely, generated as a consequence of the dynamics of the relevant theory from consistency)  associated with the breaking of the scale symmetry which is however broken anomalously.  
\par
Another advantage is that the vacuum condensate $\sigma_{vc}$ of mass dimension 2 in Yang-Mills theory  is calculable analytically \cite{Schaden99,KS00,VKAV01} and also numerically \cite{Boucaudetal01,Boucaudetal02}.
The non-vanishing vacuum condensate stems from the gluon pairing \cite{Fukuda78,GM82} at least in the Landau gauge.  In the gauges other than the Landau, we need to take into account the ghost--antighost pairing \cite{Schaden99,KS00} to maintain the BRST invariance, as pointed out in \cite{Kondo01}.    
It is also interesting to examine the confinement \cite{KondoIII} in quantum electrodynamics (QED) due to photon pairing \cite{Fukuda89,IF91} which is expected to occur in the strong coupling phase \cite{Miransky85,BLL86,KMY89,ASTW88}.
\par
Thus the dual Ginzburg-Landau theory is able to describe the dual superconductivity of the Yang-Mills vacuum.  
 The type of dual superconductor can be determined by specifying the action of $h$ more accurately.  The quadratic action in $h$ corresponds to the London limit, an extreme case of the type II, where the radial degrees of freedom of the scalar field is freezed and only the phase degrees of freedom is active.
Within the quadratic approximation, we have obtained a finite value of the string tension after removing the ultraviolet cutoff. 
  The removal of the cutoff was made possible first in this paper by taking into account the next-to-leading order of the expansion which was overlooked so far.  
\par
 To go beyond the London limit, it is necessary to specify the quartic term which is also calculable from the Yang-Mills theory as a higher-order cumulant according to the strategy indicated in this paper.  
 In order to perform the calculation, however, we need to know more detailed properties of the correlation function of the gluon field (and ghost field).  
\par
The integration of massive modes or high-energy modes can be interpreted as a step of Wilsonian RG.  
For a justification of this viewpoint, however, we need to perform more serious studies of the Wilsonian RG, as has been tried in \cite{Ellwanger98,Freire01,EW01}.
This issue will be investigated also in a subsequent paper \cite{IKKMS02}.
\par
Moreover, it is not yet clear whether the large $N$ (color) limit simplifies this project and is able to give more clear connection between the Yang-Mills theory and the string theory.    
 This issue is to be investigated in future.


\section*{Acknowledgments}
One of the authors (K.-I. K.) would like to thank Prof. Hugo Reinhardt for the kind hospitality at T\"ubingen University, Prof. Olivier Pene at LPT, Universit\'e Paris-sud (Universit\'e de Paris XI)  and Prof. Laurent Baulieu at LPTHE, Universite Pierre et Marie Curie (Universit\'e de Paris VII) where a part of this work was performed.
The authors would like to thank Irina Aref'eva
for drawing our attention to some old papers.
This work is supported in part by Sumitomo Foundations and by 
Grant-in-Aid for Scientific Research from the Ministry of
Education, Science and Culture: (B)13135203 and (C)14540243 .

\appendix
\section{Cumulant expansions}
\label{sec:cumulant}

\par
The cumulant expansion up to the order $\kappa_0^{-2}$ reads

\begin{align}
  F[h,\mathscr{A}] 
  =&    {1 \over \kappa_0} 
  \Big\langle \int d^4x  \left( {1 \over 2}h_{\mu\nu} f^V_{\mu\nu}-{1 \over 4}f^V_{\mu\nu}f^V_{\mu\nu} \right)   
   \Big\rangle^{\text{con}}_V
  \nonumber\\
  &+  {1 \over 2\kappa_0^2}  
  \Big\langle  \int d^4x \int d^4y \left( {1 \over 2}h_{\mu\nu}(x) f^V_{\mu\nu}(x)-{1 \over 4}f^V_{\mu\nu}(x)f^V_{\mu\nu}(x) \right)   
  \nonumber\\
 & \times \left( {1 \over 2}h_{\alpha\beta}(y) f^V_{\alpha\beta}(y)-{1 \over 4}f^V_{\alpha\beta}(y)f^V_{\alpha\beta}(y) \right)   
   \Big\rangle^{\text{con}}_V + \cdots .
   \label{expa}
\end{align}
The integration by parts allows us to rewrite the space-time integral of $h_{\mu\nu} f^V_{\mu\nu}$ into 
\begin{align}
   \int d^4x   {1 \over 2} h_{\mu\nu}(x) f^V_{\mu\nu}(x) 
   =& \int d^4x  [\partial_\nu h_{\mu\nu}(x)] [{\bf V}(x) \cdot \mathscr{A}_\mu(x)]
   \nonumber\\
&  - {1 \over 2}  \int d^4x \ g^{-1}h_{\mu\nu}(x) {\bf V}(x) \cdot (\partial_\mu{\bf V}(x) \times \partial_\nu{\bf V}(x)) .
\label{hf}
\end{align}
First, we consider the contribution to the double cumulant expansion from $m=1$:  
$
  g\langle   F[h,\mathscr{A}]  \rangle_{YM}^{\text{con}} .
$
In what follows, we omit all the terms which do not couple to $h_{\mu\nu}$ field.  
Hence, the first term in (\ref{expa}) reads
\begin{align}
  & {1 \over \kappa_0} \Big\langle \Big\langle \int d^4x {1 \over 2}h_{\mu\nu}(x)  f^V_{\mu\nu}(x) \Big\rangle_{V}^{\text{con}} \Big\rangle_{YM}^{\text{con}}
\nonumber\\
   =& {1 \over \kappa_0} \int d^4x [\partial_\nu h_{\mu\nu}(x)] \langle {\bf V}(x) \rangle_{V}^{\text{con}} \cdot \langle \mathscr{A}_\mu(x) \rangle_{YM}^{\text{con}}
   \nonumber\\
&  - {1 \over 2\kappa_0} \int d^4x g^{-1}h_{\mu\nu}(x) \langle {\bf V}(x) \cdot (\partial_\mu{\bf V}(x) \times \partial_\nu{\bf V}(x)) \rangle_{V}^{\text{con}} 
   \nonumber\\
=& - {1 \over 2\kappa_0}  \int d^4x g^{-1}h_{\mu\nu}(x) \langle \Omega_{\mu\nu}(x) \rangle_{V}^{\text{con}} ,
\label{1st}
\end{align}
where we have used 
$\langle {\bf V}(x) \rangle_{V}^{\text{con}} = \langle \mathscr{A}_\mu(x) \rangle_{YM}^{\text{con}} = 0$
and defined
\begin{equation}
 \Omega_{\mu\nu}(x):={\bf V}(x) \cdot (\partial_\mu{\bf V}(x) \times \partial_\nu{\bf V}(x)) .
\end{equation}

\par
Similarly, the second term in (\ref{expa}) has the expectation value:
\begin{align}
  &   {1 \over 2\kappa_0^2}  \Big\langle \Big\langle \int d^4x {1 \over 2}h_{\mu\nu}(x) f^V_{\mu\nu}(x) \int d^4y  {1 \over 2}h_{\alpha\beta}(y)  f^V_{\alpha\beta}(y) \Big\rangle_{V}^{\text{con}}
  \Big\rangle_{YM}^{\text{con}}
  \nonumber\\
 =&     {1 \over 2\kappa_0^2} \int d^4x \int d^4y 
 \partial_\nu h_{\mu\nu}(x) \partial_\beta h_{\alpha\beta}(y) 
 \langle V^A(x) V^B(y) \rangle_{V}^{\text{con}} \langle \mathscr{A}_\mu^A(x) \mathscr{A}_\alpha^B(y) \rangle_{YM}^{\text{con}} 
 \nonumber\\
 & +  {1 \over 8\kappa_0^2} g^{-2} \int d^4x \int d^4y 
h_{\mu\nu}(x) h_{\alpha\beta}(y)
\langle \Omega_{\mu\nu}(x) \Omega_{\alpha\beta}(y) \rangle_{V}^{\text{con}} , 
\end{align}
where we have used 
$\langle \mathscr{A}_\mu(x) \rangle_{YM}^{\text{con}} = 0$ again.
\par
Next, we consider the contribution from $m=2$:
$
  \langle  (  F[h,\mathscr{A}] )^2  \rangle_{YM}^{\text{con}}  .
$
By taking into account (\ref{1st}),
the contribution comes from
\begin{align}
  &  \left( \Big\langle \int d^4x  {1 \over 2}h_{\mu\nu}(x) f^V_{\mu\nu}(x) \Big\rangle_{V}^{\text{con}} \right)^2 
\nonumber\\&    
   =  {1 \over 4}  \int d^4x \int d^4y g^{-2} h_{\mu\nu}(x) h_{\alpha\beta}(y) \langle \Omega_{\mu\nu}(x) \rangle_{V}^{\text{con}} \langle \Omega_{\alpha\beta}(y) \rangle_{V}^{\text{con}}   .
\end{align}
We can put
$\langle \Omega_{\mu\nu}(x) \rangle_{V}^{\text{con}} = 0$
from Lorentz invariance.
Thus we arrive at (\ref{nlaction}).
\par

\section{Manifestly covariant quantization of antisymmetric tensor field}
\label{sec:KRfield}

In this appendix, we adopt the Minkowski formulation. 

\subsection{Massless case}

We discuss the gauge fixing of a second-rank antisymmetric tensor gauge field $A_{\mu\nu}$ whose Lagrangian in Minkowski space-time is given by 
\begin{equation}
  {\cal L}_0 = - {1 \over 8}(\epsilon_{\mu\nu\rho\sigma}\partial^\nu A^{\rho\sigma})^2 .
\end{equation}
This Lagrangian is invariant under the hypergauge transformation,
\begin{equation}
  \delta A_{\mu\nu}(x) = 
\partial_\mu \xi_\nu(x) - \partial_\nu \xi_\mu(x) .
\end{equation}
In order to fix the gauge, we adopt the gauge fixing condition for $A_{\mu\nu}$,
\begin{equation}
 \partial^\nu A_{\mu\nu} = 0 .
\end{equation}
Then the gauge fixing (GF) and Faddeev-Popov (FP) ghost term is obtained based on the prescription of Kugo and Uehara \cite{KU82} as
\begin{equation}
 {\cal L}_1 := - i\bm{\delta}_{\rm B}\left[ \bar C^\nu( \partial^\mu A_{\mu\nu} + {\alpha_1 \over 2}B_\nu) \right] ,
\end{equation}
where we have introduced the vector ghost $C$, antighost $\bar{C}$ and the Nakanishi-Lautrup (NL) field $B$.  
Note that $C$ and $\bar C$ are independent fields and that
$C^\dagger=C$, $\bar C^\dagger=\bar C$. 
Then the nil potent BRST transformations are defined by
\begin{align}
  \bm{\delta}_{\rm B} A_{\mu\nu}(x) =& 
\partial_\mu C_\nu(x) - \partial_\nu C_\mu(x) ,
\nonumber\\
  \bm{\delta}_{\rm B} C_\mu(x) =& i \partial_\mu d(x),
\nonumber\\
  \bm{\delta}_{\rm B} d(x) =& 0 ,
\nonumber\\
  \bm{\delta}_{\rm B} \bar C_\mu(x) =& i B_\mu(x) ,
\nonumber\\
  \bm{\delta}_{\rm B} B_\mu(x) =& 0 .
\end{align}
Hence, the explicit form of ${\cal L}_1$ reads
\begin{equation}
 {\cal L}_1 =  B^\nu \partial^\mu A_{\mu\nu} 
+ i \bar C^\nu \partial^\mu [\partial_\mu C_\nu - \partial_\nu C_\mu]
+ {\alpha_1 \over 2}(B_\mu)^2 .
\end{equation}
\par
However, the Lagrangian ${\cal L}_1$ and hence ${\cal L}_0+{\cal L}_1$ is still invariant under the transformation of the vector ghosts $C_\mu$ and $\bar C_\mu$, i.e., 
$\delta C_\mu(x) = i \partial_\mu \theta(x),
\delta \bar C_\mu(x) = i \partial_\mu \varphi(x)$.  
Therefore we must fix the gauge for the vector ghost and antighost.  Here we consider the gauge fixing conditions,
$\partial^\mu \bar C_\mu= 0$ and $\partial^\mu C_\mu=0$.
Thus we obtain an additional GF+FP term,
\begin{equation}
 {\cal L}_2 := - i\bm{\delta}_{\rm B}\left[ \bar d \left( \partial^\mu C_{\mu}
 + \alpha_2 P \right) \right]
- i \bm{\delta}_{\rm B} \left[ N \left( \partial^\mu \bar C_\mu +  \alpha_3 B^{(1)} \right) \right] .
\end{equation}
where the nil potent BRST transformations of the additional fields are supplemented as
\begin{align}
  \bm{\delta}_{\rm B} N(x)  =&   P(x) ,
\nonumber\\
  \bm{\delta}_{\rm B} P(x)  =& 0 ,
\nonumber\\
  \bm{\delta}_{\rm B} \bar d(x)  =&  B^{(1)}(x) ,
\nonumber\\
  \bm{\delta}_{\rm B} B^{(1)}(x)  =& 0 .
\end{align}
The explicit form of ${\cal L}_2$ reads 
\begin{equation}
 {\cal L}_2 = -i B^{(1)}\partial^\mu C_{\mu} -i  \alpha_4 B^{(1)}P
+ \bar d \partial^\mu \partial_\mu d
-i P \partial^\mu \bar C_\mu   + N \partial^\mu B_\mu  ,
\end{equation}
where we have defined $\alpha_4 :=\alpha_2-\alpha_3$.  Note that $P$ and $B^{(1)}$ anti-commute.
For the assignment of the ghost number of each field, see Table.1.
Two vector fields $C_\mu$ and $\bar C_\mu$ are primary ghosts, and three scalar fields $d$, $\bar d$ and $N$ are secondary ghosts. Three fields $B_\mu, P$ and $B^{(1)}$ are the Lagrange multiplier fields for the condition,
$\partial^\nu A_{\mu\nu}=0$, $\partial^\mu \bar C_\mu= 0$ and $\partial^\mu C_\mu=0$,
respectively.  Thus we obtain the GF+FP term for the Lagrangian ${\cal L}_0$, i.e.,
${\cal L}_{GF+FP}={\cal L}_1 + {\cal L}_2$.
Now all the gauge degrees of freedom are fixed.  Thus the full Lagrangian density is given by
\begin{align}
 {\cal L}_{tot} =& {\cal L}_0 + {\cal L}_{GF+FP} ,
\nonumber\\ 
   =&  {\cal L}_0  + B^\nu \partial^\mu A_{\mu\nu} 
+ i \bar C^\nu \partial^\mu [\partial_\mu C_\nu - \partial_\nu C_\mu]+ {\alpha_1 \over 2}(B_\mu)^2 
\nonumber\\& 
-i B^{(1)}\partial^\mu C_{\mu} -i  \alpha_4 B^{(1)}P
+ \bar d \partial^\mu \partial_\mu d
-i P \partial^\mu \bar C_\mu   + N \partial^\mu B_\mu  .
\end{align}
The massless antisymmetric tensor field stands for the massless spin-0 field as a physical mode.
It is possible to show that all the unphysical modes decouple leaving correctly one physical mode \cite{Kimura80}.
\footnote{
The above result is the summary of the results obtained by Townsend \cite{Townsend79}, Kimura \cite{Kimura80} and Hata, Kugo and Ohta \cite{HKO81}.
The same result can be obtained within the framework of the extended theory for the constrained system based on the canonical Hamiltonian formalism on the extended phase space, the so-called the Batalin-Fradkin-Vilkovisky (BFV) formalism, see e.g. the original papers and a review \cite{BFV}.
}

\begin{table}
\begin{center}
 \begin{tabular}{c||cc}
field & rank  & ghost number \\
\hline
\hline
$A$      & 2 & 0 \\
$C$      & 1 & 1 \\
$d$      & 0 & 2 \\
$\bar C$ & 1 & -1 \\
$B$      & 1 & 0 \\
$N$      & 0 & 0 \\
$P$      & 0 & 1 \\
$\bar d$ & 0 & -2 \\
$B^{(1)}$ & 0 & -1 \\
\hline
$\Lambda$ & 1 & 0 \\
$C'$      & 0 & 1 \\
$\bar C'$ & 0 & -1 \\
$B'$      & 0 & 0 
 \end{tabular}
\end{center}
\caption{The ghost number of the field with the indicated rank.}
\label{table1}
\end{table}

\subsection{Massive case}

Next, we consider the theory of an antisymmetric tensor field with the mass term.
The Lagrangian in Minkowski space-time is given by
\begin{equation}
  {\cal L}_0^m[A] 
= - {1 \over 8}(\epsilon_{\mu\nu\rho\sigma}\partial^\nu A^{\rho\sigma})^2 
 - {1 \over 4} m^2 (A_{\mu\nu})^2 .
\end{equation}
This Lagrangian with the mass term is no longer invariant under the hypergauge transformation of $A_{\mu\nu}$.  However, the invariance is recovered by introducing an additional vector field $\Lambda_\mu$ in such a way that
\begin{equation}
  {\cal L}_0^m{}'[A,\Lambda] 
= - {1 \over 8}(\epsilon_{\mu\nu\rho\sigma}\partial^\nu A^{\rho\sigma})^2 
 - {1 \over 4} (mA_{\mu\nu}+\partial_\mu \Lambda_\nu - \partial_\nu \Lambda_\mu)^2 .
\end{equation}
Actually, this Lagrangian is invariant under the combined transformation,
\begin{equation}
  \delta A_{\mu\nu}(x) = 
\partial_\mu \xi_\nu(x) - \partial_\nu \xi_\mu(x) ,
\quad \delta \Lambda_\mu(x) = - m \xi_\mu(x) .
\end{equation}
Moreover, it has another invariance under the transformation,
\begin{equation}
  \delta A_{\mu\nu}(x) = 0 ,
\quad \delta \Lambda_\mu(x) = \partial_\mu \omega(x).
\end{equation}
\par
Therefore, we define the BRST transformation of $A$ and $\Lambda$ as
\begin{align}
  \bm{\delta}_{\rm B} A_{\mu\nu}(x) =& \partial_\mu C_\nu(x) - \partial_\nu C_\mu(x) ,
\label{BRST-A}
\\
  \bm{\delta}_{\rm B} \Lambda_\mu(x) =& - m C_\mu(x) + \partial_\mu C'(x) ,
\label{BRST-Lambda}
\end{align}
where we have introduced the ghost $C$ and an extra ghost $C'$.
The BRST transformations of $C$ and $C'$ are determined by the nil potency as 
\begin{align}
  \bm{\delta}_{\rm B} C_\mu(x) =& i \partial_\mu d(x)  ,
\\
 \bm{\delta}_{\rm B} C'(x) =& im d(x), 
\\
 \bm{\delta}_{\rm B} d(x) =& 0 .
\end{align}
  The BRST transformation of antighosts  $\bar{C}$ and $\bar{C'}$ is defined by
\begin{align}
 \bm{\delta}_{\rm B} \bar{C}_\mu(x)  =& iB_\mu(x) ,
\\
 \bm{\delta}_{\rm B} B_\mu(x)  =& 0 ,
\\
 \bm{\delta}_{\rm B} \bar C'(x)  =& iB'(x) ,
\\
 \bm{\delta}_{\rm B} B'(x)  =& 0 ,
\\
 \bm{\delta}_{\rm B} \bar d(x)  =&  B^{(1)}(x) ,
\\
 \bm{\delta}_{\rm B} B^{(1)}(x)  =& 0 .
\end{align}
Moreover, we introduce $N$ and $P$ in such a way that 
\begin{align}
 \bm{\delta}_{\rm B} N(x)  =& P(x) ,
\\
 \bm{\delta}_{\rm B} P(x)  =& 0 .
\end{align}
Thus the nil potent BRST transformation is determined for all the fields.
In order to fix the gauge degrees of freedom, we must add the GF+PF term ${\cal L}_{GF+FP}$.
\par
A good choice is%
\footnote{  
The authors would like to thank Atsushi Nakamura \cite{Nakamura00} for helpful discussions on this Appendix.
}
\begin{align}
  {\cal L}_{GF+FP}  =& - i \bm{\delta}_{\rm B} \Biggr[
\bar C^\nu \left( \partial^\mu A_{\mu\nu} - \partial_\nu N - a \Lambda_\nu 
+{\alpha_1 \over 2}B_\nu  \right)
+ \bar d \left( \partial^\mu C_\mu + b C' + \alpha_2 P \right)
\nonumber\\& 
+ \bar C' \left( \partial^\mu \Lambda_\mu+{\alpha' \over 2}B' \right) \Biggr] 
\\
 =& B^\nu \left( \partial^\mu A_{\mu\nu} - \partial_\nu N - a \Lambda_\nu 
+{\alpha_1 \over 2}B_\nu  \right)
\nonumber\\& 
+ i \bar C^\nu [\partial^\mu (\partial_\mu C_\nu-\partial_\nu C_\mu) 
-\partial_\nu P-a(\partial_\nu C'-m C_\nu)]
\nonumber\\& 
-iB^{(1)}(\partial^\mu C_\mu+bC'+\alpha_2 P)
+ \bar d \partial^\mu \partial_\mu d + b m \bar d d 
\nonumber\\& 
+ B'(\partial^\mu \Lambda_\mu+{\alpha' \over 2}B')
+ i\bar C' \partial^\mu (\partial_\mu C'-mC_\mu) ,
\end{align}
where $\alpha_1, \alpha_2, \alpha{}'$ are gauge fixing parameters and $a,b$ are parameters specified later.
Roughly speaking, this corresponds to the gauge fixing condition,
$\partial^\mu A_{\mu\nu}=0$, $\partial^\mu C_\mu=0=\partial^\mu \bar C_\mu$ and
$\partial^\mu \Lambda_\mu=0$.
For the total Lagrangian ${\cal L}_{tot}' := {\cal L}_0^m{}' + {\cal L}_{GF+FP}$,
the generating functional of the theory is given by
\begin{equation}
  Z := \int {\cal D}A_{\mu\nu} {\cal D}\Lambda_\mu {\cal D}B_\mu
{\cal D}C_\mu {\cal D}\bar C_\mu {\cal D}d {\cal D}\bar d {\cal D}N 
{\cal D}P {\cal D}B^{(1)}{\cal D}C'{\cal D}\bar C' {\cal D}B'
\exp \left[ i \int d^4x {\cal L}_{tot}' \right] .
\end{equation}
\par
After integrating over $B$ and $B'$,
the sector containing $A$ and $\Lambda$ reads
\begin{equation}
  Z := \int {\cal D}A_{\mu\nu} {\cal D}\Lambda_\mu {\cal D}N  
\exp \left[ i \int d^4x {\cal L}_{tot}'' \right] ,
\end{equation}
with
\begin{align}
 {\cal L}_{tot}''  =& {\cal L}_0^m{}[A] 
- {1 \over 2}mA_{\mu\nu}(\partial_\mu \Lambda_\nu - \partial_\nu \Lambda_\mu) - {1 \over 4}  (\partial_\mu \Lambda_\nu - \partial_\nu \Lambda_\mu)^2
\nonumber\\& 
 - {1 \over 2\alpha'}(\partial^\mu \Lambda_\mu)^2 
- {1 \over 2\alpha_1}(\partial^\mu A_{\mu\nu}-\partial_\nu N-a\Lambda_\nu)^2  ,
\end{align}
since other fields decouple from the relevant sector.
Then, performing the integration over $N$, we obtain
\begin{align}
  Z  =& \int {\cal D}A_{\mu\nu} 
\exp \left\{ i \int d^4x \left( {\cal L}_0^m{}[A] 
- {1 \over 2\alpha_1} \left( \partial^\nu A_{\mu\nu} \right)^2 \right) \right\}
\nonumber\\&  \times
\int {\cal D}\Lambda_\mu
 \exp \left\{ i \int d^4x {\cal L}_1^m{}[A,\Lambda] \right\},
\end{align}
where we have defined
\begin{align}
{\cal L}_1^m{}[A,\Lambda]  :=& 
- {1 \over 2}mA_{\mu\nu}(\partial_\mu \Lambda_\nu - \partial_\nu \Lambda_\mu) - {1 \over 4}  (\partial_\mu \Lambda_\nu - \partial_\nu \Lambda_\mu)^2
\nonumber\\
&-   {a^2 \over 2\alpha_1}(\Lambda_\mu)^2
+ {a \over \alpha_1}(\partial^\mu A_{\mu\nu})\Lambda_\nu 
+ {a^2 \over 2\alpha_1}(\partial^\mu \Lambda_\mu)\Delta^{-1}(\partial^\nu \Lambda_\nu)  .
\end{align}
If we choose $a=m\alpha_1$, the cross term $(\partial^\nu A_{\mu\nu})\Lambda_\mu$ cancels with 
$mA_{\mu\nu}(\partial_\mu \Lambda_\nu - \partial_\nu \Lambda_\mu)$.
Hence, ${\cal L}_1^m{}[A,\Lambda]$ becomes independent of $A$ field:
\begin{equation}
{\cal L}_1^m{}[\Lambda] =
 - {1 \over 4}m^2 (\partial_\mu \Lambda_\nu - \partial_\nu \Lambda_\mu)^2
+  {\alpha_1 m^2 \over 2}(\Lambda_\mu)^2
+ {\alpha_1 m^2 \over 2}(\partial^\mu \Lambda_\mu)\Delta^{-1}(\partial^\nu \Lambda_\nu)  .
\end{equation}
Thus the $\Lambda$ field decouples from the theory of $A$.  
Consequently, the theory with 
${\cal L}_{tot}'[A,\Lambda] := {\cal L}_0^m{}' + {\cal L}_{GF+FP}$ 
is equivalent to the theory with 
$
{\cal L}_1[A] := {\cal L}_0^m[A] - {1 \over 2\alpha_1} \left( \partial^\nu A_{\mu\nu} \right)^2
$.  In particular, the choice of the Landau gauge $\alpha_1=0$ yields
\begin{equation}
  Z = \int {\cal D}A_{\mu\nu}\delta(\partial^\nu A_{\mu\nu}) 
\exp \left\{ i \int d^4x  {\cal L}_0^m{}[A]   \right\} .
\label{Th1}
\end{equation}
\par
On the other hand, if we choose $a=0$, the theory with 
${\cal L}_{tot}'[A,\Lambda] := {\cal L}_0^m{}' + {\cal L}_{GF+FP}$ 
is equivalent to the theory with 
$
{\cal L}_2[A] := {\cal L}_0^m{}'[A,\Lambda] - {1 \over 2\alpha_1} \left( \partial^\nu A_{\mu\nu} \right)^2
$.
In particular, the choice of the Landau gauge $\alpha_1=0$ yields
\begin{equation}
  Z = \int {\cal D}A_{\mu\nu} {\cal D}\Lambda_\mu  \delta(\partial^\nu A_{\mu\nu}) 
\exp \left\{ i \int d^4x  {\cal L}_0^m{}'[A,\Lambda]   \right\} .
\label{Th2}
\end{equation}
Note that the massive antisymmetric tensor gauge theory stands for the massive spin-1 theory \cite{Kawai81,SS01}.
The antisymmetric tensor $A_{\mu\nu}=- A_{\nu\mu}$ has six components in four-dimensional space-time.  
The gauge fixing condition $\partial^\nu A_{\mu\nu}=0$ yields three independent relations among the six components, since a trivial condition $\partial^\mu \partial^\nu A_{\mu\nu}=0$ is satisfied for the antisymmetric tensor $A_{\mu\nu}$.  Therefore, $A_{\mu\nu}$ has three independent components.  This number agrees with the degrees of freedom for the massive vector field of spin one.  
\par
Moreover, it is possible to consider a simpler gauge \cite{BS96},
\begin{align}
  {\cal L}_{GF+FP}  =& - i \bm{\delta}_{\rm B} \left[
\bar C^\mu \Lambda_\mu + \bar d C' \right] 
\\
 =&  B^\mu \Lambda_\mu + i \bar C^\mu (\partial_\mu C' - mC_\mu)
-iB^{(1)}C' + m\bar d d .
\end{align}
This corresponds to the gauge fixing condition,
$\Lambda_\mu=0$ and $C'=0$.
The integration over $N$, $P$ and $B'$ is trivial, since ${\cal L}_{tot}'$ does not include them.  The $B$ integration leads to the constraint 
$\delta(\Lambda_\mu)$,
\begin{equation}
  Z = \int {\cal D}A_{\mu\nu} {\cal D}\Lambda_\mu \delta(\Lambda_\mu)
\int {\cal D}C_\mu {\cal D}\bar C_\mu {\cal D}d {\cal D}\bar d {\cal D}B^{(1)} {\cal D}C'{\cal D}\bar C'
\exp \left[ i \int d^4x {\cal L}_{tot}'' \right] ,
\end{equation}
where
\begin{equation}
 {\cal L}_{tot}'' = {\cal L}_0^m{}'[A,\Lambda] 
  + i \bar C^\mu (\partial_\mu C' - mC_\mu)
-iB^{(1)}C' + m\bar d d    .
\end{equation}
When we consider the sector of $A$ and $\Lambda$, the sector described by other fields decouples and we obtain

\begin{equation}
  Z = \int {\cal D}A_{\mu\nu} {\cal D}\Lambda_\mu \delta(\Lambda_\mu)
\exp \left\{ i \int d^4x {\cal L}_0^m{}'[A,\Lambda] \right\} 
= \int {\cal D}A_{\mu\nu} 
\exp \left\{ i \int d^4x {\cal L}_0^m[A] \right\} . 
\label{Th3}
\end{equation}
Therefore, we recover the original theory which is written by the $A$ field only with the Lagrangian ${\cal L}_0^m$.

\section{Integration over antisymmetric tensor field and the boundary term}
\label{sec:Binteg}

By including the source term, the Lagrangian is given by
\begin{align}
 \mathcal{L}_d[B] =& {1 \over 4\kappa} B_{\mu\nu}B_{\mu\nu}
+ {1 \over 4\eta^2} 
(-B_{\mu\nu} \partial^2 B_{\mu\nu} - B_{\mu\nu} \partial_\nu \partial_\lambda B_{\lambda\mu} + B_{\nu\mu} \partial_\mu \partial_\lambda B_{\lambda\nu})
\nonumber\\&
- {1 \over 4\kappa^2} (-B_{\mu\nu} B_{\mu\nu} - B_{\mu\nu} \partial_\nu {1 \over \partial^2} \partial_\lambda B_{\lambda\mu} + B_{\nu\mu} \partial_\mu {1 \over \partial^2} \partial_\lambda B_{\lambda\nu})
- {1 \over 4\omega^2} B_{\mu\nu} I_{\mu\nu\rho\sigma} \partial^2 B_{\rho\sigma} 
\nonumber\\&
- {1 \over 4\gamma^4} [-B_{\mu\nu} (\partial^2)^2 B_{\mu\nu} - B_{\mu\nu} \partial^2 \partial_\nu \partial_\lambda B_{\lambda\mu} + B_{\nu\mu} \partial^2 \partial_\mu \partial_\lambda B_{\lambda\nu}] 
+ i {1 \over 2}B_{\mu\nu} \mathcal{J}_{\mu\nu} ,
\label{Bcomp}
\end{align}
where 
$\mathcal{J}_{\mu\nu}:=gJ \ {}^*\Theta_{\mu\nu}$.
The saddle point equation of the Lagrangian in the momentum   representation is given by
\begin{align}
  & {1 \over \kappa} \tilde B_{\mu\nu}(p)
+ \left( {1 \over \eta^2} - {1 \over \kappa^2} {1 \over p^2} \right)
[ p^2 \tilde B_{\mu\nu}(p) + p_\nu p_\lambda \tilde B_{\lambda\mu}(p) - p_\mu p_\lambda \tilde B_{\lambda\nu}(p)]
\nonumber\\&
+ {p^2 \over \omega^2}  I_{\mu\nu\rho\sigma} \tilde B_{\rho\sigma}(p) 
 + {1 \over \gamma^4} p^2 
[ p^2 \tilde B_{\mu\nu}(p) + p_\nu p_\lambda \tilde B_{\lambda\mu}(p) - p_\mu p_\lambda \tilde B_{\lambda\nu}(p)] 
+ i  \tilde{\mathcal{J}}_{\mu\nu}(p) = 0 .
\end{align}
\par
Here we introduce the projection operators $I$ and $P$ defined by
\begin{align}
  I_{\mu\nu\rho\sigma} :=& {1 \over 2}(\delta_{\mu\rho} \delta_{\nu\sigma} - \delta_{\mu\sigma} \delta_{\nu\rho}) ,
\\
 P_{\mu\nu\alpha\beta} :=& {1 \over 2}(T_{\mu\alpha}T_{\nu\beta}-T_{\mu\beta}T_{\nu\alpha}), \quad 
T_{\mu\nu} := \delta_{\mu\nu} - {p_\mu p_\nu \over p^2} ,
\end{align}
where
\begin{align}
  I_{\mu\nu\rho\sigma} I_{\rho\sigma\alpha\beta}
 = I_{\mu\nu\alpha\beta}, 
\quad
  P_{\mu\nu\rho\sigma} P_{\rho\sigma\alpha\beta} 
= P_{\mu\nu\alpha\beta} ,
\\
 I_{\mu\nu\rho\sigma} P_{\rho\sigma\alpha\beta}
= P_{\mu\nu\rho\sigma} I_{\rho\sigma\alpha\beta} 
 = P_{\mu\nu\alpha\beta} .
\end{align}
Note that the projection operators have the properties: 
\begin{align}
 I_{\mu\nu\alpha\beta} =& - I_{\nu\mu\alpha\beta} = I_{\alpha\beta\mu\nu}
= - I_{\mu\nu\beta\alpha} ,
\\
 P_{\mu\nu\alpha\beta} =& - P_{\nu\mu\alpha\beta} = P_{\alpha\beta\mu\nu}
= - P_{\mu\nu\beta\alpha} ,
\end{align}
and
\begin{align}
p^{\rho_1} I_{\mu\nu\rho_1\sigma} p^{\rho_2} I_{\alpha\beta\rho_2\sigma}
=&  {1 \over 4}(p_\mu p_\alpha g_{\nu\beta}-p_\mu p_\beta g_{\nu\alpha}
-p_\nu p_\alpha g_{\mu\beta}+p_\nu p_\beta g_{\mu\alpha})
\nonumber\\
 =& {1 \over 2}p^2 (I-P)_{\mu\nu\alpha\beta} .
 \label{I-P}
\end{align}
For an arbitrary antisymmetric tensor $A_{\mu\nu}$, it follows that
\begin{align}
 I_{\mu\nu\alpha\beta} A^{\alpha\beta} = A_{\mu\nu},
\quad
 A^{\mu\nu}I_{\mu\nu\alpha\beta} = A_{\alpha\beta} ,
\\
 p^2 P_{\mu\nu\alpha\beta} A^{\alpha\beta} 
= p^2 A_{\mu\nu} - p_\nu p_\beta A_{\mu\beta} + p_\mu p_\beta A_{\nu\beta} .
\end{align}
\par
Thus the saddle point equation is rewritten as
\begin{equation}
  \left[ \left( {1 \over \kappa} + {p^2 \over \omega^2} \right) I_{\mu\nu\alpha\beta} + \left( -{1 \over \kappa^2} + {p^2 \over \eta^2}  + {p^4 \over \gamma^4} \right) P_{\mu\nu\alpha\beta} \right] \tilde B_{\alpha\beta}(p) = - i  \tilde{\mathcal{J}}_{\mu\nu}(p) .
\end{equation}
It is shown using the properties of the projection operators that the inverse of the operator $[aI+bP]$ is  
$(a+b)^{-1}[ I+(b/a)(I-P)]$.  Hence the saddle point value of $B$ is 
\begin{align}
  \tilde B_{\mu\nu}(p) 
  =  { \tilde{\kappa} M_1^2 M_2^2 \over (p^2+M_1^2)(p^2+M_2^2)} 
\left[ I_{\mu\nu\alpha\beta} + 
{ {-1 \over \kappa} + {p^2 \over M^2} (1  + {p^2 \over \tilde M^2})  
\over 1  + {p^2 \over M'{}^2} }  
(I-P)_{\mu\nu\alpha\beta} \right]
 (- i \tilde{\mathcal{J}}_{\alpha\beta}(p)) ,
\end{align}
where we have defined 
$M^2:=\eta^2/\kappa$, $\tilde M^2 := \gamma^4/\eta^2$, $M'{}^2 := \omega^2/\kappa$,
$\tilde \kappa :=(1/\kappa + 1/\kappa^2)^{-1}$, 
 and ($\gamma^4=\tilde{\kappa} M_1^2 M_2^2$) 
\begin{align}
 (M_{2,1})^2 := {1 \over 2}\gamma^4 \left( {1 \over \eta^2}+{1 \over \omega^2} \right) \left[ 1 \pm \sqrt{1-{4 \over \tilde{\kappa}}\gamma^{-4} \left( {1 \over \eta^2}+{1 \over \omega^2} \right)^{-2}} \right] .
\end{align}
By substituting this value back into the Lagrangian in the momentum representation, the $B$ integration is performed to obtain
\begin{align}
& \exp \Biggr\{ - {\tilde{\kappa} \over 4} \int {d^4 p \over (2\pi)^4} e^{ip(x-y)}   \tilde{\mathcal{J}}_{\mu\nu}(-p)  \tilde \Sigma_{\alpha\beta}(p)
 \left( {\chi \over p^2+M_1^2} - {\chi \over p^2+M_2^2} \right)
 \nonumber\\&
 \quad \quad \times 
 \left[ I_{\mu\nu\alpha\beta} + { {-1 \over \kappa} + {p^2 \over M^2} (1  + {p^2 \over \tilde M^2})  
\over 1  + {p^2 \over M'{}^2} }   
(I-P)_{\mu\nu\alpha\beta} \right]
 \Biggr\} .
\end{align}
If we neglect the term proportional to $(I-P)$, this agrees with the argument of the exponential of (\ref{W}). 
It turns out that
the same result is obtained by the action (\ref{Bth2}) which is obtained from (\ref{Bcomp}) by setting $\partial_\lambda B_{\lambda\mu}=0$.
The London limit is obtained by taking the limit
$\gamma, \omega \rightarrow \infty$.
\par
It is possible to show that the term proportional to $(I-P)$ yields the boundary term.  The identity (\ref{I-P}) implies 
\begin{align}
& \exp \Biggr\{ - {\tilde{\kappa} \over 4} \int d^4x   \mathcal{J}_{\mu\nu}(x)  \int d^4y \Sigma_{\alpha\beta}(y)
\int {d^4 p \over (2\pi)^4} e^{ip(x-y)} 
 \left( {\chi \over p^2+M_1^2} - {\chi \over p^2+M_2^2} \right)
 \nonumber\\&
 \quad \quad \times 
 \left[ I_{\mu\nu\alpha\beta} + { {-1 \over \kappa} + {p^2 \over M^2} \left(1  + {p^2 \over \tilde M^2} \right)  
\over 1  + {p^2 \over M'{}^2} }   
(I-P)_{\mu\nu\alpha\beta} \right]
 \Biggr\}
 \nonumber\\ 
 =& \exp \Biggr\{ - {\tilde{\kappa} \over 4} \int d^4x   \mathcal{J}_{\mu\nu}(x)  \int d^4y \Sigma_{\alpha\beta}(y)
 \Biggr[
I_{\mu\nu\alpha\beta} \int {d^4 p \over (2\pi)^4}
 \left( {\chi \over p^2+M_1^2} - {\chi \over p^2+M_2^2} \right) e^{ip(x-y)}
 \nonumber\\&
 \quad \quad  
 +2 g_{\nu\beta}\partial_\mu^x \partial_\alpha^y H(x,y)   \Biggr]
 \Biggr\} ,
\end{align}
where
\begin{equation}
 H(x,y) := 
 \int {d^4 p \over (2\pi)^4}
 \left( {\chi \over p^2+M_1^2} - {\chi \over p^2+M_2^2} \right)
 {1 \over p^2}
   { {-1 \over \kappa} + {p^2 \over M^2} \left(1  + {p^2 \over \tilde M^2} \right)  \over 1  + {p^2 \over M'{}^2} }   
 e^{ip(x-y)} .
\end{equation}
Hence the second term in the exponential is written using the partial integration into the boundary term:
\begin{align}
   -{\tilde{\kappa} \over 2} \int d^4x  \partial_\mu^x \mathcal{J}_{\mu\nu}(x)  \int d^4y  \partial_\alpha^y  \Sigma_{\alpha\nu}(y) H(x,y) 
  =  -{\tilde{\kappa} \over 2} g^2J^2  \oint_C dx_\nu \oint_C dy_\nu H(x,y) .
\end{align}

\section{Calculation of the Wilson loop average}
\label{sec:Wilsonloop}

We choose a rectangular loop with side lengths $R$ and $T$ in the $x_1-x_4$ plane:
\begin{equation}
  \Theta_{\mu\nu}(z) = \delta_{\mu 1}\delta_{\nu 4} \delta(z_2) \delta(z_3) \theta(z_1) \theta(R-z_1) \theta(z_4) \theta(T-z_4) .
\end{equation}
Then the Fourier transformation is obtained as
\begin{align}
  \Theta_{\mu\nu}(p)  \equiv& \int d^4z \Theta_{\mu\nu}(z) e^{-ip \cdot z}
\nonumber\\ 
 =& \delta_{\mu 1}\delta_{\nu 4} \int_0^R dz_1 e^{-ip_1 z_1}
\int_0^T dz_4 e^{-ip_4 z_4}
\nonumber\\ 
 =& \delta_{\mu 1}\delta_{\nu 4} {2 \over p_1} e^{-i{p_1 R \over 2}} \sin {p_1 R \over 2} {2 \over p_4} e^{-i{p_4 T \over 2}} \sin {p_4 T \over 2} .
\label{Theta1}
\end{align}
In the momentum representation, we have
\begin{align}
(\Theta, I(-\partial^2+M_i^2)^{-1} \Theta) 
= \int {d^4p \over (2\pi)^4}  \Theta_{\mu\nu}(p)
 I_{\mu\nu\alpha\beta} (-\partial^2+M_i^2)^{-1}(p) \Theta_{\alpha\beta}(-p).
\label{nonl}
\end{align}
 For large $L=R$ and $L=T$, we can apply to (\ref{Theta1}) the formula:
\begin{equation}
  \lim_{L \rightarrow \infty}\left({\sin aL \over a}\right)^2
 = \pi L \delta(a) ,
\end{equation}
  Then (\ref{nonl}) reads 
\begin{align}
(\Theta, I(-\partial^2+M_i^2)^{-1} \Theta) 
 \cong&  \int {d^4p \over (2\pi)^4} (2\pi)^2 T R \delta(p_1)\delta(p_4) 
 (-\partial^2+M_i^2)^{-1}(p)
\nonumber\\
 =& T R  \int {d^2p \over (2\pi)^2}   
(-\partial^2+M_i^2)^{-1}(0,p_2,p_3,0)
\nonumber\\
 =&  T R  \int {d^2p \over (2\pi)^2}  
{1 \over p_2^2+p_3^2+M_i^2}   \quad (i=1,2) .
\end{align}
Each integral is logarithmically divergent.
By taking into account two terms, the logarithmic divergence of the integral is removed after introducing an ultraviolet cutoff $\Lambda$ as 
\begin{align}
  \sum_{i=1,2} (-1)^{i+1} (\Theta, I(-\partial^2+M_i^2)^{-1} \Theta) 
 =&  T R   \int {d^2p \over (2\pi)^2}   \left[ 
{1 \over p_2^2+p_3^2+M_1^2} - {1 \over p_2^2+p_3^2+M_2^2} \right] 
\nonumber\\
 =&  TR   \lim_{\Lambda \rightarrow \infty} \int_0^{\Lambda^2} {d |p|^2 \over 4\pi}  \left[ 
{1 \over |p|^2+M_1^2} - {1 \over |p|^2+M_2^2} \right]
\nonumber\\
 =& TR   {1 \over 4\pi} \lim_{\Lambda \rightarrow \infty} \ln {\Lambda^2+M_1^2 \over M_1^2}{M_2^2 \over \Lambda^2+M_2^2}
\nonumber\\
 =& TR {1 \over 2\pi} \ln {M_2 \over M_1} .
\label{areacontribution}
\end{align}
\par
Similarly, we can calculate the inner product:
\begin{align}
  (\Theta, I \Theta)
:=& \int d^4x \Theta_{\mu\nu}(x) I_{\mu\nu\rho\sigma} \Theta_{\rho\sigma}(x) 
= \int {d^4p \over (2\pi)^4} \Theta_{\mu\nu}(p) \Theta_{\mu\nu}(-p) 
\nonumber\\
 \cong&  \int {d^4p \over (2\pi)^4} (2\pi)^2 T R \delta(p_1)\delta(p_4) 
\nonumber\\
 =& TR  \int {d^2p \over (2\pi)^2}  
= TR {\Lambda^2 \over 4\pi} .
\end{align}
This part corresponds to the quadratic divergence (i.e., the volume divergence in two-dimensional space).  This should be removed by the  renormalization prescription.  
\par
Therefore we obtain the $\Lambda$-independent and finite result.
Thus the Wilson loop average exhibits the area law decay,
\begin{equation}
  W(C) \cong \exp ( - \sigma_{st} RT ) ,
  \quad 
  \sigma_{st} = J^2 g^2 \tilde{\kappa}  {\chi \over 8\pi} \ln {M_2 \over M_1} .
\end{equation}

\section{Derivative expansion of the non-local string action}
\label{sec:dexp}

In this section, we begins with the expression,
\begin{align}
   W(C) 
=&  \exp \left[ -  {1 \over 4} J^2g^2 \tilde{\kappa}
  \int_{\Sigma_C} dS^{\mu\nu}(x) \int_{\Sigma_C} dS^{\rho\sigma}(y) 
   G_{\mu\nu\rho\sigma}(x,y) \right] ,
\\
   G_{\mu\nu\rho\sigma}(x,y) 
 :=&    I_{\mu\nu\rho\sigma}
  \left( {\chi \over - \Delta+M_1^2} - {\chi \over -\Delta+M_2^2} \right) ,
\end{align}
where
$
  \chi :=  M_1^2 M_2^2/(M_2^2-M_1^2).
$
We define the Euclidean propagator:
\begin{equation}
  G_M(x) = (-\Delta_E+M^2)^{-1}(x,0)
= \int {d^4k \over (2\pi)^4} e^{ik \cdot x} {1 \over k^2+M^2} .
\label{Gm}
\end{equation}
The propagator is obtained in the closed form: 
\begin{equation}
  G_M(x) = {1 \over 4\pi^2}{M \over |x|} K_1(M|x|)
=  {1 \over 4\pi^2} M^2 {K_1(|x|/\xi) \over |x|/\xi},
\label{Dfunc}
\end{equation}
where $K_1(z)$ is the modified Bessel function and $\xi$ is the correlation length defined by $\xi=M^{-1}$.
In fact, (\ref{Dfunc})  is obtained as follows.
Substituting the identity,
\begin{equation}
  {1 \over k^2+M^2} = \int_0^\infty ds e^{-s(k^2+M^2)} ,
\end{equation}
into (\ref{Gm}) and performing the Gaussian integration over the four momenta $k$, we obtain
\begin{align}
  G_M(x)  =& \int_0^\infty ds e^{-sM^2} 
\int {d^4k \over (2\pi)^4} e^{- sk^2+ik \cdot x }  
\nonumber\\
 =&  \int_0^\infty ds e^{-sM^2} 
\exp \left[ -{x^2 \over 4s} \right] {1 \over (2\pi)^4}
\left( \sqrt{{\pi \over s}} \right)^{4}
\nonumber\\
 =& {1 \over 16\pi^2} \int_0^\infty ds {1 \over s^2}
\exp \left[ -sM^2 -{x^2 \over 4s} \right] .
\end{align}
The above result (\ref{Dfunc}) is immediately obtained by applying the integration formula\cite{GR80}:
\begin{equation}
  \int_0^\infty ds \ s^{\nu-1} \exp \left[ -{\beta \over s} -M^2 s 
\right] 
= 2 (\beta/ M^2)^{\nu/2} K_\nu(2\sqrt{\beta M^2}) 
\quad (\Re \beta > 0, \Re M^2 > 0)),
\end{equation}
to the case $\nu=-1$ and $\beta = x^2/4$, since 
$K_{-\nu}(z)=K_{\nu}(z)$.
\par
In Euclidean space,
\begin{equation}
 G_{\mu\nu\rho\sigma}(x,x') 
=     I_{\mu\nu\rho\sigma} \chi
  \left[ G_{M_1}(x-x') - G_{M_2}(x-x') \right]  .
\label{F}
\end{equation}
We define
\begin{equation}
   J_i :=  \int_{S_C} dS^{\mu\nu}(x(\sigma)) \int_{S_C} dS^{\rho\sigma}(x(\sigma{}')) 
   I_{\mu\nu\rho\sigma} G_{M_i}((x(\sigma)-x(\sigma{}'))^2) .
\end{equation}
It is shown \cite{AES96,Antonov99} that the derivative expansion of $J_i$ in powers of 
\begin{equation}
  \zeta_i^a := (\sigma{}'-\sigma)^a/\xi_i ,
\quad \xi_i := M_i^{-1} ,
\end{equation}
leads to
\begin{equation}
   J_i =  \int d^2\sigma \sqrt{g} \left[ 4 \xi_i^2 \mathscr{M}_i^{(0)}
- {1 \over 4}\xi_i^4 \mathscr{M}_i^{(2)} g^{ab}(\partial_a t_{\mu\nu})(\partial_b t_{\mu\nu}) \right] 
+ O(\xi_i^6 ) ,
\end{equation}
where no summation is understood over $i$ and we have defined the moment,
\begin{equation}
  \mathscr{M}_i^{(n)} := \int d^2z (z_i^{2})^n G_{M_i}(z^2) , \quad z_i^a := g^{1/4}\zeta_i^a .
\end{equation}
Here we have used the conformal gauge for the induced metric, 
$g_{ab}(\sigma)=\sqrt{g(\sigma})\delta_{ab}$ which leads to 
$\zeta^a \zeta^b g_{ab}=g^{-1/2}g_{ab}z^a z^b=z^a z^b \delta_{ab}:=z^2$.
Thus, the confining string theory derived in this paper is characterized by the parameters,
\begin{subequations}
\begin{align}
  \sigma_{st}  =& J^2g^2 \tilde{\kappa} \chi \int d^2z [M_1^{-2}G_{M_1}(z^2)-M_2^{-2}G_{M_2}(z^2)] ,
\label{stension}
\\
 \alpha_0^{-1}  =& - {1 \over 16} J^2g^2 \tilde{\kappa} \chi \int d^2z z^2 [M_1^{-4} G_{M_1}(z^2) - M_2^{-4} G_{M_2}(z^2)] ,
\label{alphaeq}
\\
 \kappa_t =& {1 \over 24} J^2g^2 \tilde{\kappa} \chi \int d^2z z^2 [M_1^{-4} G_{M_1}(z^2) - M_2^{-4} G_{M_2}(z)] .
\end{align}
\end{subequations}
By substituting (\ref{Dfunc}) into (\ref{stension}), we obtain
\begin{align}
  \sigma_{st} 
 =&  J^2g^2 \tilde{\kappa} {\chi \over 4\pi^2} \left[ 
\int_{{M_1 \over \Lambda}}^{\infty} 2\pi |z| d|z|  {K_1(|z|) \over |z|}
- \int_{{M_2 \over \Lambda}}^{\infty} 2\pi |z| d|z| {K_1(|z|) \over |z|} \right] 
\nonumber\\
 =& J^2g^2 \tilde{\kappa} {\chi \over 2\pi} \left[ K_0\left({M_1 \over \Lambda} \right) - K_0\left({M_2 \over \Lambda} \right) \right]  ,
\end{align}
where we have used $K_1(x) = - K_0'(x)$ and $K_0(\infty)=0$ and introduced an ultraviolet cutoff $\Lambda$ (since $K_0(0)=\infty$). 
Note that the asymptotic behavior of the modified Bessel function $K_0(z)$ for $z \ll 1$ is given by
\begin{equation}
 K_0(z) \cong - (\gamma_E + \ln {z \over 2}) = \ln {2e^{-\gamma_E} \over z} ,
\end{equation}
with $\gamma_E$ being Euler's constant $\gamma_E=0.5772\cdots$.
Thus, for sufficiently large $\Lambda$, we obtain the $\Lambda$-independent {\it finite} result.%
\footnote{The string tension is a free energy per unit length of the string.  It is well known that the free energy has a logarithmic dependence in the Ginzburg-Landau theory.  
The London limit corresponds to $M_2 \rightarrow \infty$.  Hence, the string tension reduces to the expression,
$
\sigma_{st} = J^2g^2 \tilde{\kappa} {\chi \over 2\pi} K_0 \left( {M_1 \over \Lambda} \right) ,
$
since
$K_0(M_2/\Lambda) \rightarrow 0$.
}
 Incidentally, the asymptotics of $K_p(z)$ for $z>0$,
\begin{equation}
 K_p(z) \sim \sqrt{{\pi \over 2z}} e^{-z} [1+O(z^{-1})] ,
\end{equation}
means that $K_p(z)$ decreases exponentially for large $z$.
After removing the cutoff $\Lambda$, the above expression reduces to a finite value,
\begin{equation}
 \sigma_{st}  
\cong {J^2g^2  \over 2\pi} \tilde{\kappa} \chi \ln \left( {M_2 \over M_1} \right) .
\end{equation}
This is an advantage of our approach over the previous one \cite{AE99}.    
The coefficient of the rigidity term is calculated as
\begin{align}
 \alpha_0^{-1}  =& -{1 \over 16} J^2g^2 \tilde{\kappa} {\chi \over 4\pi^2} \int_{0}^{\infty} 2\pi |z| d|z| \left[ {|z|^2 \over M_1^2}{K_1(|z|) \over |z|}
- \int_{0}^{\infty} |z| d|z| {|z|^2 \over M_2^2}{K_1(|z|) \over |z|} \right] 
\nonumber\\
 =& - J^2g^2 \tilde{\kappa} {\chi \over 64\pi^2} \left[ {4\pi \over M_1^2} 
-   {4\pi \over M_2^2}  \right]  
\nonumber\\
 =& - J^2g^2 \tilde{\kappa} {1 \over 16\pi}  <0  ,
\label{alpha0}
\end{align}
where we have used the integration formula ($\nu=1, \mu=3$),
\begin{equation}
 \int_0^\infty dx x^{\mu-1} K_\nu(ax) = 2^{\mu-2}a^{-\mu} \Gamma\left({\mu-\nu \over 2}\right) \Gamma\left({\mu+\nu \over 2}\right) \quad (\Re \mu > \Re \nu) .
\end{equation}
Here note that the integral in (\ref{alpha0}) is finite and we don't have to introduce the cutoff.
Similarly, the parameter $\kappa_t$ is calculated as
\begin{equation}
 \kappa_t = -{2 \over 3} \alpha_0^{-1} 
 = {2 \over 3} J^2g^2 \tilde{\kappa} {1 \over 16\pi}  > 0  .
\end{equation}

\section{Path-integral duality transformations}
\label{sec:dual}

\subsection{Dual Ginzburg-Landau theory in the London limit}

\par
We introduce the field strength $H$ of an antisymmetric tensor field $B$ (the so-called Kalb-Ramond field) by $H:=dB$, i.e., 
\begin{equation}
 H_{\mu\nu\lambda} := \partial_\lambda B_{\mu\nu} + \partial_\mu B_{\nu\lambda}
+ \partial_\nu B_{\lambda\mu}
\end{equation}
Note that the field strength $H$ is invariant, 
$H_{\mu\nu\lambda}  \rightarrow H_{\mu\nu\lambda} $,
under the hypergauge transformation, $B \rightarrow B+d\zeta$, i.e.,
\begin{equation}
 B_{\mu\nu} \rightarrow B_{\mu\nu}^\zeta := B_{\mu\nu} + \partial_\mu \zeta_\nu  - \partial_\nu \zeta_\mu  .
\label{gtr1}
\end{equation}
We require the invariance of the measure ${\cal D}B_{\mu\nu}$ under the hypergauge transformation.  
However,  the Lagrangian ${\cal L}_d[B]$ does not have the invariance under the hypergauge transformation due to the existence of the mass term $(B_{\mu\nu})^2$.  
Nevertheless, we can recover the hypergauge invariance of the antisymmetric tensor theory by introducing a new vector field%
\footnote{This field plays the similar role to the St\"uckelberg scalar field in the massive vector theory which recovers the gauge invariance of the vector field.}
 $\Lambda_\mu$ which transforms as 
\begin{equation}
 \Lambda_\mu \rightarrow \Lambda_\mu^\zeta := \Lambda_\mu - \zeta_\mu .
\label{gtr2}
\end{equation}
In fact, the combination $B^\Lambda:=B+d\Lambda$, i.e., 
\begin{equation}
 B_{\mu\nu}^\Lambda = B_{\mu\nu} + \partial_\mu \Lambda_\nu  - \partial_\nu \Lambda_\mu ,
\end{equation}
is invariant under the combined transformations, (\ref{gtr1}) and (\ref{gtr2}).
Therefore, the Lagrangian,
\begin{equation}
{\cal L}_m[B,\Lambda] :=  {\cal L}_m[B^\Lambda] =  {1 \over 4\kappa} (B_{\mu\nu}+ \partial_\mu \Lambda_\nu  - \partial_\nu \Lambda_\mu)^2
  + {1 \over 12\eta^2} (H_{\mu\nu\lambda})^2 
+ {1 \over 4\gamma^4} (\partial^\lambda H_{\lambda\mu\nu})^2  ,
\label{magtheory}
\end{equation}
is also invariant, i.e.,
${\cal L}_m[B,\Lambda]={\cal L}_m[B^\zeta, \Lambda^\zeta]$.

 As we have recovered the hypergauge invariance, we need to fix the hypergauge invariance in quantizing the dual magnetic theory ${\cal L}_m[B,\Lambda]$.  
For this purpose, we adopt the gauge-fixing condition,%
\begin{equation}
\partial^\nu B_{\mu\nu} =0  ,
\label{Bgfc}
\end{equation}
for the antisymmetric tensor field.%
\footnote{In the manifestly covariant quantization of the gauge theory, we need to introduce the ghost as is well known.  However, it is not enough for the antisymmetric tensor gauge theory, since we need to introduce the ghost for ghost in order to completely fix the gauge degrees of freedom.  Such a theory is called a reducible theory. In this subsection we treat the theory in a naive manner.  However, the result is unchanged if we take into account the reducibility of the theory.  See Appendix~\ref{sec:KRfield}.} 
Under this  condition,
 the derivative terms of $B_{\mu\nu}$ in ${\cal L}_m[B^\Lambda]$ reproduce the corresponding terms of ${\cal L}_d[B]$,
\begin{align}
(H_{\mu\nu\lambda} )^2 =& - 3 B_{\mu\nu} \partial^2 B_{\mu\nu} 
-6 \partial_\mu B_{\mu\nu} \partial_\lambda B^{\nu\lambda} 
\rightarrow 
- 3 B_{\mu\nu} \partial_\lambda \partial^\lambda B^{\mu\nu}  ,
\nonumber\\
(\partial^\lambda H_{\lambda\mu\nu} )^2  
=& (\partial^\lambda \partial_\lambda B_{\mu\nu} + \partial_\mu \partial^\lambda B_{\nu\lambda}+\partial_\nu \partial^\lambda B_{\lambda\mu})^2
\rightarrow 
  B_{\mu\nu} (\partial_\lambda \partial^\lambda)^2 B^{\mu\nu}  .
\end{align}
Therefore, ${\cal L}_m[B,\Lambda]$ under the condition (\ref{Bgfc}) reduces to 
\begin{equation}
{\cal L}_m[B,\Lambda] = {1 \over 4\kappa} (B_{\mu\nu} + \partial_\mu \Lambda_\nu - \partial_\nu \Lambda_\mu )^2
  - {1 \over 4\eta^2} B_{\mu\nu} \partial_\lambda \partial^\lambda B^{\mu\nu} 
+ {1 \over 4\gamma^4} B_{\mu\nu} (\partial_\lambda \partial^\lambda)^2 B^{\mu\nu}  .
\end{equation}
It is shown that the theory with ${\cal L}_m[B,\Lambda]$ reduces to the original theory given by ${\cal L}_d[B]$ by integrating out $\Lambda$ field after fixing the gauge freedom of $\Lambda_\mu$, see Appendix~\ref{sec:KRfield}. 
\par
Thus we obtain an alternative dual description of low-energy Gluodynamics in terms of $B_{\mu\nu}$ and $\Lambda_\mu$. Especially, for $G=SU(2)$, we obtain
\begin{equation}
    W(C) 
= Z_{M}^{-1}   \int  {\cal D}B_{\mu\nu}  
  \delta(\partial^\nu B_{\mu\nu}) \int  {\cal D}\Lambda_\mu
\exp \left\{ - S_{M}[B^\Lambda;C]  \right\}  ,
\end{equation}
where
\begin{equation}
S_{M}[B^\Lambda;C] =  \int d^4x {\cal L}_m[B^\Lambda]
+ {i \over 2} Jg \int_{S_C} dS^{\mu\nu} B_{\mu\nu} .
\label{Mth}
\end{equation}
Apart from the third term in ${\cal L}_m[B,\Lambda]$, the above action (\ref{magtheory}) coincides with the action of confining string proposed by Polyakov \cite{Polyakov96} in the weak field limit.

We define
\begin{equation}
S_{M}[B^\Lambda;\Theta] =  \int d^4x \left\{ {\cal L}_m[B^\Lambda] 
+ {i \over 2} Jg  B_{\mu\nu}  \ {}^*\Theta_{\mu\nu} \right\} .
\label{APEGTh}
\end{equation}
Then the Wilson loop average is given by
\begin{align}
    W(C) 
= Z_{M}[B^\Lambda;\Theta]/Z_{M}[B^\Lambda;0],
\label{vevh}
\end{align}
where
\begin{align}
 & Z_{M}[B^\Lambda;\Theta] 
\nonumber\\
 :=& 
\int  {\cal D}B_{\mu\nu}   \delta(\partial^\nu B_{\mu\nu}) \int {\cal D}\zeta_\mu 
\exp \left\{ - S_{M}[B^\zeta;\Theta]  \right\}  .
\nonumber\\
 =& 
 \int  {\cal D}B_{\mu\nu}  \delta(\partial^\nu B_{\mu\nu})
\exp \left\{ - \int d^4x \left[ {1 \over 12 \eta^2} (H_{\lambda\mu\nu})^2
+ {1 \over 4\gamma^4} (\partial^\lambda H_{\lambda\mu\nu})^2 \right] \right\}
\nonumber\\
 & \times \exp \left\{ -i \int d^4x  {1 \over 2} Jg B_{\mu\nu} \ {}^*\Theta_{\mu\nu} \right\}
\nonumber\\
 & \times \int {\cal D}\zeta_\mu 
\exp \left\{ -  \int d^4x  {1 \over 4\kappa} (B_{\mu\nu}^\zeta)^2  \right\} .
\end{align}
Note that the path integral transformation holds,
\begin{align}
 &  \int {\cal D}\zeta_\mu 
\exp \left\{ -  \int d^4x  {1 \over 4\kappa} (B_{\mu\nu}^\zeta
)^2  \right\}
\nonumber\\
 =&
  \int {\cal D}\ell_{\mu\nu} \delta(\epsilon^{\mu\nu\rho\sigma}\partial_\rho
(\ell_{\mu\nu}-B_{\mu\nu} ))
\exp \left\{ -  \int d^4x  {1 \over 4\kappa} (\ell_{\mu\nu})^2  \right\} ,
\label{ell}
\end{align}
since the constraint, 
\begin{equation}
 \epsilon^{\mu\nu\rho\sigma}\partial_\rho
(\ell_{\mu\nu}-B_{\mu\nu} ) = 0 ,
\end{equation}
is solved by 
\begin{equation}
  \ell_{\mu\nu}-B_{\mu\nu} 
= \partial_\mu \zeta_\nu- \partial_\nu \zeta_\mu ,
\quad i.e, \quad
  \ell_{\mu\nu} = B_{\mu\nu}^{\zeta} .
\end{equation}
\par
Moreover, we introduce the auxiliary (Abelian) vector field $U_\mu$ by
\begin{equation}
  \delta(\epsilon^{\mu\nu\rho\sigma}\partial_\rho
(\ell_{\mu\nu}-B_{\mu\nu} 
))
= \int {\cal D}U_\mu \exp \left\{-  i \int d^4x    \ {}^*U_{\mu\nu}
 (\ell_{\mu\nu}-B_{\mu\nu} 
)  \right\} ,
\label{iden}
\end{equation}
where $U_{\mu\nu}$ is the (dual) field strength defined by
\begin{equation}
 U_{\mu\nu}:=\partial_\mu U_\nu - \partial_\nu U_\mu .
\end{equation}
By using the identity (\ref{iden}),  the integration over $\ell_{\mu\nu}$ in (\ref{ell}) can be performed as
\begin{align}
 &  (\ref{ell})
\nonumber\\
 =& \int {\cal D}U_\mu \exp \left\{ i \int d^4x    \ {}^*U_{\mu\nu}  B_{\mu\nu}   \right\}
\int {\cal D}\ell_{\mu\nu} \exp \left\{ -  \int d^4x  \left[ 
{1 \over 4\kappa} (\ell_{\mu\nu})^2 + i \ {}^*U_{\mu\nu} \ell_{\mu\nu} 
\right] \right\}
\nonumber\\
 =& \int {\cal D}U_\mu \exp \left\{ - \int d^4x  \left[ 
  \kappa (U_{\mu\nu})^2 - i \ {}^*U_{\mu\nu} B_{\mu\nu} \right] \right\} .
\end{align}
Hence the equality holds,%
\footnote{
Another way of deriving the equality is as follows.  
The argument of the exponential in the LHS is
\begin{equation}
 (B+d\zeta, B+d\zeta) = (B,B) + (B,d\zeta) + (d\zeta, B) + (d\zeta,d\zeta)
\sim (B,B) + (d\zeta,d\zeta),
\end{equation}
 under the condition $\delta B=0$.
The last term decouples after the Gaussian integration of $\zeta$.
On the other hand, the argument of the exponential in the RHS is cast into
\begin{equation}
 \int d^4x  \left[  (U_{\mu\nu})^2 - \ {}^*U_{\mu\nu} (B_{\mu\nu} 
) \right] 
= (dU,dU) - (*dU,B) 
= (U,\delta dU) - (U,*dB) .
\end{equation}
Suppose the Lorentz type gauge condition $\delta U=0$. We introduce the NL  (zero-form) field $\phi$. Then the Gaussian integration over $U_\mu$ field yields
\begin{align}
  & (U,\Delta U)-(U,*dB) - (\delta U,\phi)
= (U,\Delta U)-(U,*dB+d\phi)  
\nonumber\\
 &\rightarrow 
(*dB+d\phi,{1 \over \Delta} *dB+d\phi)
= (B, {\delta d \over \Delta}B) + (\phi, {\delta d \over \Delta}\phi)
\sim (B,B) + (\phi,\phi),
\end{align}
under the condition $\delta B=0$.  In this derivation, we must insert the constraint $\delta(\partial^\mu U_\mu)$ in the measure ${\cal D}U_\mu$.
The identity implies that there are many ways of extracting the transverse modes of $B$.
}
\begin{equation}
 \int {\cal D}\zeta_\mu 
\exp \left\{ -  \int d^4x  {1 \over 4\kappa} (B_{\mu\nu}^\zeta)^2  \right\}
=  \int {\cal D}U_\mu \exp \left\{ - \int d^4x  \left[ 
  \kappa (U_{\mu\nu})^2 -i \ {}^*U_{\mu\nu} B_{\mu\nu} \right] \right\} .
\end{equation}
Thus, the theory is rewritten in terms of $U_\mu$ and $B_{\mu\nu}$ as 
\begin{align}
 & Z_{M}[U,B;\Theta] 
\nonumber\\
 =&  \int {\cal D}U_\mu
\exp \left\{ - \int d^4x \kappa (U_{\mu\nu})^2 \right\}
\int  {\cal D}B_{\mu\nu}  \delta(\partial^\nu B_{\mu\nu})
\nonumber\\&  \times
\exp \left\{ - \int d^4x \left[ 
  - i \ {}^*U_{\mu\nu} B_{\mu\nu}  
+ {1 \over 12 \eta^2} (H_{\lambda\mu\nu})^2 
+ {1 \over 4\gamma^4} (\partial^\lambda H_{\lambda\mu\nu} )^2
\right] \right\}  
\nonumber\\
 & \times \exp \left\{ -i \int d^4x  {1 \over 2}Jg   B_{\mu\nu} \ {}^*\Theta_{\mu\nu} \right\} .
\end{align}
By change of variable 
$U_{\mu\nu} \rightarrow U_{\mu\nu} + {1 \over 2}Jg \Theta_{\mu\nu}$, we arrive at the expression,
\begin{align}
 &  Z_{M}[U,B;\Theta] 
\nonumber\\
 =&  \int {\cal D}U_\mu
\exp \left\{ - \int d^4x   \kappa \left( U_{\mu\nu} + {1 \over 2}Jg \Theta_{\mu\nu} \right)^2 \right\}
\nonumber\\&  \times
\int  {\cal D}B_{\mu\nu}  \delta(\partial^\nu B_{\mu\nu})
\exp \left\{ - \int d^4x \left[ 
  -i \ {}^*U_{\mu\nu} B_{\mu\nu}  
+ {1 \over 12 \eta^2} (H_{\lambda\mu\nu})^2 
+ {1 \over 4\gamma^4} (\partial^\lambda H_{\lambda\mu\nu} )^2
\right] \right\} .
\label{Last}
\end{align}

\par

\par
We change the variable $B_{\mu\nu}$ into the new variable $Y_\mu$ as%
\footnote{
From $\delta B^{(2)}=0$, there exists a three-form $Z^{(3)}$ such that 
$B^{(2)}=\delta Z^{(3)}=\delta *W^{(1)}=*dW^{(1)}$.  Then 
$H^{(3)}:=dB^{(2)}=d*dW^{(1)}=*\delta dW^{(1)}=*Y^{(1)}$, or
$Y^{(1)}=*H^{(3)}$.  Therefore, 
$\delta Y^{(1)}=\delta *H^{(3)}=*dH^{(3)}=*ddB^{(2)}=0.$
}
\begin{align}
 &  \int  {\cal D}B_{\mu\nu}  \delta(\partial^\nu B_{\mu\nu})
\exp \left\{ - \int d^4x \left[ 
- i \ {}^*U_{\mu\nu} B_{\mu\nu} 
+ {1 \over 12 \eta^2} (H_{\lambda\mu\nu})^2 
+ {1 \over 4 \gamma^4} (\partial^\lambda H_{\lambda\mu\nu} )^2
 \right] \right\}
\nonumber\\
 =& \int  {\cal D}B_{\mu\nu}  \delta(\partial^\nu B_{\mu\nu})
\exp \left\{ - \int d^4x \left[
- i  \epsilon^{\mu\nu\rho\sigma} U_{\mu} \partial_\nu B_{\rho\sigma}  + {1 \over 12 \eta^2} (H_{\lambda\mu\nu})^2 
+ {1 \over 4 \gamma^4} (\partial^\lambda H_{\lambda\mu\nu} )^2
\right] \right\}
\nonumber\\
 =& \int {\cal D}Y_\mu \delta(\partial_\mu Y^\mu) \exp \left\{ - \int d^4x \left[
- i  2U_\mu Y^\mu  
+ {1 \over 2\eta^2}Y_\mu^2 
+ {1 \over 4 \gamma^4}(\partial_\mu Y_\nu - \partial_\nu Y_\mu)^2  
\right] \right\} ,
\label{int}
\end{align}
since the constraint $\partial_\mu Y^\mu=0$ can be solved by an antisymmetric tensor field in the form,
\begin{equation}
 Y^\mu := {1 \over 2} \epsilon^{\mu\nu\rho\sigma} \partial_\nu B_{\rho\sigma}  .
\label{constr2}
\end{equation}
The massive antisymmetric tensor field $B_{\mu\nu}$ denotes the massive spin-1 field $Y_\mu$ whose canonical mass dimension is three. 
This should be compared with the massless antisymmetric tensor field which stands for the massless spin-0 field, see \cite{Townsend79,Kimura80,HKO81}.
In this step, the number of independent degrees of freedom is conserved, since $Y_\mu$ and $B_{\mu\nu}$ have three independent components.
The path-integral duality transformation indicates the correspondence:
\begin{equation}
 Y^\mu \leftrightarrow {1 \over 2} \epsilon^{\mu\nu\rho\sigma} \partial_\nu B_{\rho\sigma} 
= {1 \over 6} \epsilon^{\mu\nu\rho\sigma} H_{\nu\rho\sigma} 
= \partial_\nu \ {}^*B^{\mu\nu} 
\leftrightarrow  \partial_\nu h^{\mu\nu}.
\label{translation}
\end{equation}

\par
Furthermore, the integration over $Y_\mu$ is performed after introducing the new variable $\theta$ to remove the delta function of the constraint $\partial_\mu Y^\mu=0$,
\begin{align}
 (\ref{int})
 =& \int {\cal D}Y_\mu \int {\cal D}\theta  \exp \left\{ 
 - \int d^4x \left[
  i \theta  \partial_\mu Y^\mu 
- 2iY^\mu U_\mu 
+ {1 \over 2\eta^2}Y_\mu^2 
+ {1 \over 4 \gamma^4}(\partial_\mu Y_\nu - \partial_\nu Y_\mu)^2  
\right] \right\}
\nonumber\\
 =& \int {\cal D}Y_\mu \int {\cal D}\theta  \exp \left\{ - \int d^4x 
 \left[
- i  Y^\mu (2U_\mu + \partial_\mu \theta)  
+ {1 \over 2\eta^2}Y_\mu^2 
+ {1 \over 4 \gamma^4}(\partial_\mu Y_\nu - \partial_\nu Y_\mu)^2  
\right] \right\}
\nonumber\\
 =&  \int {\cal D}\theta \exp \left\{ - \int d^4x \left[
{1 \over 2}(2U_\mu + \partial_\mu \theta)
{\gamma^4  \over -\partial^2 +\gamma^4/\eta^2} \left( g^{\mu\nu} - {\eta^2 \over \gamma^4} \partial^\mu \partial^\nu \right) 
(2U_\nu + \partial_\nu \theta)
\right] \right\} .
\end{align}

\par
Finally, after rescaling of $U_\mu \rightarrow {1 \over 2}U_\mu$, we obtain the dual Abelian gauge theory,
\begin{align}
  & Z_{M}[U,\theta;\Theta] 
\nonumber\\
 =&    \int {\cal D}U_\mu \delta(\partial^\mu U_\mu)
 \int {\cal D}\theta
\exp \Biggr\{ - \int d^4x  \Biggr[  {\kappa \over 4} (U_{\mu\nu}+ U_{\mu\nu}^S)^2  
\nonumber\\&
\quad \quad \quad \quad \quad + 
{1 \over 2}(U_\mu + \partial_\mu \theta)
{\gamma^4  \over -\partial^2 +\gamma^4/\eta^2} \left( g^{\mu\nu} - {\eta^2 \over \gamma^4} \partial^\mu \partial^\nu \right) 
(U_\nu + \partial_\nu \theta)
\Biggr] \Biggr\} ,
\label{dAGT}
\end{align}
where  $U_{\mu\nu}^S$ is the Dirac string tensor,
\begin{align}
  U_{\mu\nu}^S(x) :=&  Jg \Theta_{\mu\nu}(x), 
\\
 \partial^\nu  U_{\mu\nu}^S(x) =&   J_\mu^S , 
\quad J_\mu^S  :=   J g \int_0^1 d\tau {dx_\mu(\tau) \over d\tau} \delta^4(x-x(\tau)) .
\end{align}
Here we have inserted the delta function $\delta(\partial^\mu U_\mu)$ for fixing the gauge for $U_\mu$.  
This model has the dual U(1) symmetry, say the magnetic $U(1)_m$ symmetry, 
\begin{equation}
 U_\mu \rightarrow U_\mu + \partial_\mu \vartheta, \quad
\theta \rightarrow \theta - \vartheta .
\end{equation}
\par
For large $\gamma$, the second term has the expansion:
\begin{align}
  & {1 \over 2} \eta^2
 (U_\mu + \partial_\mu \theta) 
\left( 1- {\eta^2 \over \gamma^4} \partial^2 \right)^{-1} \left( g^{\mu\nu} - {\eta^2 \over \gamma^4} \partial^\mu \partial^\nu \right) 
 (U_\nu + \partial_\nu \theta)
\nonumber\\
 \sim& {1 \over 2} \eta^2
  U_\mu  
\left( 1+ {\eta^2 \over \gamma^4} \partial^2  + \cdots \right)
 U^\mu  
 + {1 \over 2} \eta^2 (\partial_\mu \theta)^2 + O(\partial^4)  
\nonumber\\
 =& {1 \over 2} \eta^2 U_\mu U^\mu
 + {1 \over 2} {\eta^4 \over \gamma^4} U_\mu \partial^2 U^\mu  
 + {1 \over 2} \eta^2 (\partial_\mu \theta)^2 + O(\partial^4) ,
\label{dGL3}
\end{align}
where we have used $\partial^\mu U_\mu=0$. 
Thus we arrive at an effective theory of the Yang-Mills theory,
\begin{align}
  Z_{APEGT}[U,\theta;0] 
 =& \int {\cal D}U_\mu \delta(\partial_\mu U_\mu)   {\cal D}\theta
\exp \Biggr\{ - \int d^4x {\cal L}_{K}[U,\theta]  \Biggr\} ,
\\
 {\cal L}_{K}[U,\theta] :=&  {\kappa \over 4} \left( 1+ {\eta^4 \over \kappa \gamma^4} \right) (U_{\mu\nu})^2 
+ {1 \over 2}\eta^2 U_\mu U^\mu 
+ {1 \over 2} \eta^2 (\partial_\mu \theta)^2 .
\label{dGL4}
\end{align}

\par
After putting $\tilde \kappa=1$, this model (\ref{dGL4}) just obtained should be compared with the London limit $\lambda \rightarrow \infty$ of the dual Abelian Higgs model or the dual Ginzburg-Landau theory with the Lagrangian,
\begin{equation}
 {\cal L}_{DGL}[U,\phi] = {1 \over 4}(U_{\mu\nu}+U_{\mu\nu}^S)^2 
+ {1 \over 2}|(\partial_\mu - i2g_m U_\mu) \phi|^2
+  \lambda (|\phi|^2-v^2)^2 
\end{equation}
where $g_m$ is the magnetic charge subject to the Dirac quantization condition,
\begin{equation}
  g_m g = 4\pi .
\end{equation}
The London limit is equivalent to putting 
$|\phi(x)|=v=\text{const.}$, i.e., $\phi(x)=v \exp[i2g_m\theta(x)]$:
\begin{equation}
 {\cal L}_{DGL}[U] = {1 \over 4}(U_{\mu\nu}+U_{\mu\nu}^S)^2 
+ {1 \over 2}m_U^2 (U_\mu + \partial_\mu \theta)(U_\mu + \partial_\mu \theta). 
\end{equation}
 
\par
The effective theory (\ref{dGL4}) is of the same form as the London limit of the DGL theory, except for the renormalization of the kinetic term of the dual gauge field.
The dual gauge field becomes massive, whereas the $\theta$ field remains massless.  This is reasonable, since the field $\theta$ corresponds to the Nambu-Goldstone (NG) mode associated with the spontaneous breakdown of the magnetic U(1) symmetry.

The dual U(1) symmetry is broken in the London limit.  
This corresponds to the infinitesimally thin flux tube connecting the quark and anti-quark.  In the London limit, the Higgs mass 
$m_\phi=2\sqrt{\lambda}v$ diverges, i.e., $m_\phi=\infty$ or $m_\phi^{-1}=0$.  This is the extreme case of the type II superconductor where $m_\phi>m_U$.  
It turns out that the mass $m_U$ of the dual gauge field $U_\mu$ is given by $\eta$,
\begin{equation}
 \eta = m_U = 2g_m v \equiv {8\pi \over g}v .
\end{equation}

\subsection{Beyond the London limit}

We consider a small variation from the London limit.  
First, we introduce the fluctuation field $\varphi(x)$ around the absolute minimum $|\phi(x)|=v$:
$
 \phi(x) = v + \tau \varphi(x) ,
$ 
where $\tau := 1/\lambda \ll 1$.  
Therefore, for $\phi(x)=|\phi(x)|e^{i\theta(x)}$, the covariant derivative reads
\begin{equation}
 D_\mu \phi := \partial_\mu \phi - 2ig_m U_\mu \phi 
 = [\tau \partial_\mu \varphi+i(v+\tau \varphi)(\partial_\mu \theta - 2g_m U_\mu)] e^{i\theta} ,
\end{equation}
which leads to the kinetic term:
\begin{align}
 {1 \over 2}|D_\mu \phi|^2 
=& {1 \over 2} \tau^2 (\partial_\mu \varphi)^2 + {1 \over 2} (v+\tau \varphi)^2(\partial_\mu \theta - 2g_m U_\mu)^2 
 \nonumber\\ 
=& {1 \over 2} \tau^2 (\partial_\mu \varphi)^2 + {1 \over 2}(v^2 + 2 \tau v \varphi + \tau^2 \varphi^2 )(\partial_\mu \theta - 2g_m U_\mu)^2 .
\end{align}
Moreover, the potential reads
\begin{align}
 \lambda (|\phi|-v^2)^2 =& \lambda (|\phi|-v)^2(|\phi|+v)^2
 = \lambda \tau^2 \varphi^2 (2\eta+\tau \varphi)^2 
 \nonumber\\ 
=& 4\tau v^2 \varphi^2 + 4v \tau^2 \varphi^3 + \tau^3 \varphi^4 .
\end{align}
Hence the DGL Lagrangian up to the order $\varphi^2$ reads
\begin{align}
 {\cal L}_{DGL}[U,\phi] =& {1 \over 4}(U_{\mu\nu}+U_{\mu\nu}^S)^2 
 + {1 \over 2}v^2 (\partial_\mu \theta - 2g_m U_\mu)^2  
 \nonumber\\&
 + {\tau^2 \over 2} (\partial_\mu \varphi)^2 +   \tau  v (\partial_\mu \theta - 2g_m U_\mu)^2  \varphi
+ 4\tau v^2 \varphi^2 + O(\varphi^3) ,
\end{align}
Integrating out the fluctuation field $\varphi$, we obtain 
\begin{align}
 {\cal L}_{DGL}[U,\theta] =& {1 \over 4}(U_{\mu\nu}+U_{\mu\nu}^S)^2 
 + {1 \over 2}v^2 (\partial_\mu \theta - 2g_m U_\mu)^2  
 \nonumber\\&
 - {v^2 \over 2} (\partial_\mu \theta - 2g_m U_\mu)^2
 (-\partial^2+m^2)^{-1}  (\partial_\mu \theta - 2g_m U_\mu)^2 ,
 \label{DGL2}
\end{align}
where $m^2:=8v^2/\tau = 8\lambda v^2 \gg 1$
and $\eta:=2g_m v$.
\par
We can determine the antisymmetric tensor field theory whose dual version agrees with (\ref{DGL2}). 
  From the correspondence (\ref{translation}) and a relation (\ref{constr2}), 
\begin{equation}
  U_\mu \leftrightarrow {1 \over 2i}\left[ {1 \over \eta^2} \delta_{\mu\nu} + {1 \over \gamma^4} (\delta_{\mu\nu} \partial^2 - \partial_\mu \partial_\nu) \right] Y_\nu ,
\end{equation}
the effective theory must have an additional term
\begin{align}
 & - {1 \over 72\eta^2} \int d^4x (H_{\lambda\mu\nu})^2  (-\partial^2+m^2)^{-1} (H_{\lambda\mu\nu})^2 
 \nonumber\\
 =& - {m^{-2} \over 72\eta^2} \int d^4x (H_{\lambda\mu\nu})^2  \left(1-{\partial^2 \over m^2} \right)^{-1} \int d^4x (H_{\lambda\mu\nu})^2 
 \nonumber\\
 =& - {m^{-2} \over 72\eta^2} \int d^4x (H_{\lambda\mu\nu})^2  \left(1+{\partial^2 \over m^2} + {\partial^4 \over m^4} + \cdots \right) (H_{\lambda\mu\nu})^2 .
 \label{addh}
\end{align}
This term  is quartic in $h$ with the fourth derivative and higher derivatives. 
Therefore, the next order term in the cumulant expansion 
\begin{equation}
 - {\kappa_0^{-4} \over 4!} 
  \Big\langle
 \Big\langle \left[  \int d^4x  {1 \over 2}h_{\mu\nu} f^V_{\mu\nu} 
  - \int d^4x {1 \over 4}f^V_{\mu\nu}f^V_{\mu\nu}   
  \right]^4 \Big\rangle^{\text{con}}_V \Big\rangle^{\text{con}}_{YM} 
\end{equation}
is expected  to generate the term (\ref{addh}).
In the tree approximation for the correlation of $V$, it is easy to see that the next order term involves the piece:
\begin{align}
  &- {C^2 \over 24 \kappa_0^4} \int d^4x \int d^4y 
\langle \mathscr{A}_\mu^A(x) \mathscr{A}_\rho^B(y) \rangle_{YM} 
\langle \mathscr{A}_\nu^A(x) \mathscr{A}_\sigma^B(y) \rangle_{YM} 
\nonumber\\ 
& \times
(\partial \cdot h)_\mu(x) 
(\partial \cdot h)_\nu(x) 
(\partial \cdot h)_\rho(y) 
(\partial \cdot h)_\sigma(y) 
\nonumber\\ 
=& - {C^2 \over 24 \kappa_0^4} {1 \over 16} \int d^4x \int d^4y 
(\langle \mathscr{A}_\mu^A(x) \mathscr{A}_\mu^B(y) \rangle_{YM})^2 
(\partial \cdot h)_\rho(x) 
(\partial \cdot h)_\rho(y) 
(\partial \cdot h)_\sigma(x) 
(\partial \cdot h)_\sigma(y) 
\nonumber\\ 
=& - {C^2 \over 24 \kappa_0^4} {1 \over 16} 
\Big[ s^{(0)} 
\int d^4X 
(\partial \cdot h)^2(X) 
(\partial \cdot h)^2(X) 
\nonumber\\&
\quad \quad \quad \quad \quad
+ {1 \over 16} s^{(1)} \int d^4X 
(\partial \cdot h)^2(X) 
 \partial^2 
(\partial \cdot h)^2(X)  + \cdots \Big] ,
\end{align}
where we have defined 
$(\partial \cdot h)_\mu := \partial_\nu h_{\mu\nu}$
and the coefficient $s^{(\ell)}$ is defined by
\begin{align}
 s^{(\ell)} := \int d^4r   
(\langle \mathscr{A}_\mu^A(x) \mathscr{A}_\mu^B(y) \rangle_{YM})^2 
(r^2)^\ell \quad (\ell = 0, 1, \cdots) .
\end{align}
Indeed, this expansion reproduces the desired terms (\ref{addh}).

\par
The detailed analysis of higher order terms will be given in a forthcoming paper. Moreover, an argument supporting the above result will be given there from the viewpoint of the exact Wilsonian RG \cite{Ellwanger98}.

\end{document}